\documentclass[aps,prd,superscriptaddress,showpacs,preprintnumbers]{revtex4}
\usepackage{graphicx,color}
\newcommand{\be}{\begin{equation}}
\newcommand{\ee}{\end{equation}}
\newcommand{\bea}{\begin{eqnarray}}
\newcommand{\eea}{\end{eqnarray}}

\newcommand{\bfk}{\mbox{\boldmath $k$}}

\def\kt{k_\perp}

\newcommand{\bfq}{\mbox{\boldmath $q$}}
\newcommand{\bfP}{\mbox{\boldmath $P$}}
\newcommand{\bfS}{\mbox{\boldmath $S$}}

\newcommand{\bfz}{\mbox{\boldmath $z$}}

\newcommand{\pup}{p^\uparrow}

\newcommand{\Adown}{A^\downarrow}

\newcommand{\Aup}{A^\uparrow}

\def\lsim{\mathrel{\rlap{\lower4pt\hbox{\hskip1pt$\sim$}}\raise1pt\hbox{$<$}}}
\def\gsim{\mathrel{\rlap{\lower4pt\hbox{\hskip1pt$\sim$}}\raise1pt\hbox{$>$}}}
\def\nostrocostruttino#1\over#2{\mathrel{\mathop{\kern 0pt \rlap
{\hbox{$#1$}}} \hbox{\kern-.135em $#2$}}}

\begin{document}
\title{Sivers Effect in Drell-Yan processes}

\author{M.~Anselmino}
\affiliation{Dipartimento di Fisica Teorica, Universit\`a di Torino and
             INFN, Sezione di Torino, \\ Via P.~Giuria 1, I-10125 Torino, Italy}
\author{M.~Boglione}
\affiliation{Dipartimento di Fisica Teorica, Universit\`a di Torino and
             INFN, Sezione di Torino, \\ Via P.~Giuria 1, I-10125 Torino, Italy}
\author{U.~D'Alesio}
\affiliation{Dipartimento di Fisica, Universit\`a di Cagliari,
             I-09042 Monserrato (CA), Italy}
\affiliation{INFN, Sezione di Cagliari,
             C.P. 170, I-09042 Monserrato (CA), Italy}
\author{S.~Melis}
\affiliation{Dipartimento di Fisica Teorica, Universit\`a di Torino and
             INFN, Sezione di Torino, \\ Via P. Giuria 1, I-10125 Torino, Italy}
\author{F.~Murgia}
\affiliation{INFN, Sezione di Cagliari,
             C.P. 170, I-09042 Monserrato (CA), Italy}
\author{A.~Prokudin}
\affiliation{Dipartimento di Fisica Teorica, Universit\`a di Torino and
             INFN, Sezione di Torino, \\ Via P. Giuria 1, I-10125 Torino, Italy}
\date{\today}

\begin{abstract}
The Sivers distributions recently extracted from semi-inclusive deep inelastic
scattering data \cite{Anselmino:2008sg} are used to compute estimates
for Sivers asymmetries in Drell-Yan processes which are being planned at
several facilities (RHIC, COMPASS, J-PARC, PAX, PANDA, NICA (JINR) and SPASCHARM
(IHEP)). Most of these asymmetries turn out to be large and could allow a
clear test of the predicted sign change of the Sivers distributions when active
in SIDIS and Drell-Yan processes. This is regarded as a fundamental test of
our understanding, within QCD and the factorization scheme, of single spin
asymmetries.
\end{abstract}


\maketitle

\section{\label{Intro} Introduction}

The experimental study and theoretical understanding of transverse Single Spin
Asymmetries (SSA) has been, and still is, one of the most challenging issues
in high energy hadron physics. The original, widespread opinion that hadronic
SSAs should be negligible in any high energy process, due to the simple,
helicity conserving pQCD and Standard Model elementary dynamics
\cite{Kane:1978nd}, has been proven wrong in a great number of cases, actually
in most of the spin measurements so far performed. It often happens that the
same approach, based on the collinear QCD factorization scheme at leading twist,
which successfully describes unpolarized data, cannot explain the large spin
effects observed in the same kinematical regions.

The attempts of explaining the data and reconciling them with pQCD dynamics
have produced a much deeper understanding of the nucleon structure, involving
partonic intrinsic motion, and much progress in clarifying the mechanism, at
the partonic level, responsible for providing the phases and helicity flips
necessary for a transverse SSA. Many issues, like the universality of these
mechanisms and their correct insertion into a factorized scheme, are still
under debate and need further investigation.

We consider here the so-called Sivers asymmetry
\cite{Sivers:1989cc,Sivers:1990fh}, related to the intrinsic motion of partons
inside a transversely polarized proton, according to the distribution
\bea
\hat f_ {q/\pup} (x,\bfk_\perp) &=& f_ {q/p} (x,\kt) +
\frac{1}{2} \, \Delta^N \! f_ {q/\pup}(x,\kt)  \;
{\bfS} \cdot (\hat {\bfP}  \times
\hat{\bfk}_\perp) \label{sivnoi} \\
&=& f_ {q/p} (x,\kt) - \frac{2\,k_\perp}{m_p} \>
f_{1T}^{\perp q}(x, k_\perp) \;
{\bfS} \cdot (\hat {\bfP}  \times \hat{\bfk}_\perp) \>,
\eea
which gives the number density of unpolarized quarks $q$ (or gluons) with
intrinsic transverse momentum $\bfk_\perp$ 
inside a transversely polarized
proton $\pup$, with three-momentum $\bfP$ 
and spin polarization vector $\bfS$;
$\Delta^N \! f_ {q/\pup}(x,\kt)$, or $f_{1T}^{\perp q}(x, k_\perp)$
following a different common notation~\cite{Bacchetta:2004jz}, is the Sivers
function. A coupling of the transverse motion of unpolarized quarks and gluons
to the nucleon spin can only be related to their orbital motion, and one expects
that the observation of the Sivers asymmetry signals partonic orbital angular
momentum~\cite{Sivers:2006rg,Burkardt:2005km}. Thus, the Sivers distribution is
of very special interest.

A clear observation of a non zero Sivers distribution has been obtained by
the HERMES collaboration in Semi-Inclusive Deep Inelastic Scattering processes
(SIDIS)~\cite{Airapetian:2004tw,Diefenthaler:2007rj}. A similar measurement
by the COMPASS collaboration has given a result compatible with
zero~\cite{Alexakhin:2005iw,Martin:2007au,Alekseev:2008dn}; however, such a result
was obtained on a polarized deuteron target --- rather than a hydrogen one, as
for HERMES --- and the small result could be explained as a cancellation between
opposite contributions from $u$ and $d$ quarks. That has allowed the extraction,
from data, of the Sivers distribution function~\cite{Anselmino:2005nn,
Anselmino:2005ea,Vogelsang:2005cs, Collins:2005ie, Anselmino:2008sg}.
Large SSAs observed in $p^\uparrow p \to \pi,K + X$ inclusive processes at
Fermilab \cite{Adams:1991rx,Adams:1991cs,Adams:1995gg} and RHIC
\cite{Adams:2003fx,:2008mi} could also be explained as a manifestation of the
Sivers effect \cite{Anselmino:1994tv,Anselmino:1998yz,D'Alesio:2004up}.
A recent preliminary result from COMPASS operating on a transversely polarized
proton target \cite{Levorato:2008tv} and still compatible with zero, might, if
confirmed, be of difficult interpretation.

{From} the theoretical point of view the Sivers distribution function
\cite{Sivers:1989cc,Sivers:1990fh} had a difficult early life. For some time
its very existence was much debated and, despite its phenomenological success
\cite{Anselmino:1994tv,Anselmino:1998yz}, fundamental parity and time-reversal
properties of QCD seemed to forbid it \cite{Collins:1992kk}. Some attempts
were made of invoking initial state interactions \cite{Anselmino:1998yz} or
unusual time-reversal properties \cite{Anselmino:2001vn}. An explicit model
calculation of the Sivers effect in SIDIS \cite{Brodsky:2002cx} much clarified
the situation, showing the crucial role of final state interactions.
The analysis of Ref.~\cite{Collins:1992kk} was then reconsidered~\cite{Collins:2002kn}
and the proof that the Sivers asymmetry should vanish because of time-reversal invariance
was shown to be invalidated by the path-ordered exponential of the gluon field
in the gauge invariant operator definition of parton densities. Thus, the gauge
links (gluon exchange in final state interactions) seem to allow a
consistent QCD picture of the Sivers effect.
The same analysis \cite{Collins:2002kn} which led to the conclusion that the Sivers
distribution has a full right to existence in QCD, also showed that
time-reversal properties imply that the Sivers asymmetry must be
{\it reversed in sign} when acting in SIDIS and Drell-Yan (DY) processes.
This was explicitly confirmed by the extension of the model for Sivers effect
in SIDIS to the Drell-Yan case \cite{Brodsky:2002rv}.

The experimental measurement of the Sivers SSAs in SIDIS and Drell-Yan
processes, and the observation of the sign change of the Sivers distribution,
is one of the most important tests of our understanding of the origin
of SSAs in QCD and the factorization scheme. In this paper we exploit the
Sivers functions recently extracted from SIDIS data \cite{Anselmino:2008sg},
{\it change their signs}, and give estimates for Sivers SSAs in Drell-Yan
processes; these asymmetries can be measured in several
experiments, to be performed at existing facilities (COMPASS, RHIC),
developing (J-PARC) or proposed ones (PAX, PANDA, JINR, IHEP).

An early discussion of the contribution of the Sivers function to SSAs in
Drell-Yan processes, for RHIC experiments, was presented in
Ref.~\cite{Anselmino:2002pd}; information on the Sivers distributions was
obtained from data on SSAs in inclusive hadronic production, $\pup p \to \pi X$,
rather then from SIDIS data. This paper updates and extends subsequent studies
performed in Refs.~\cite{Vogelsang:2005cs, Anselmino:2005ea} and
\cite{Collins:2005rq}. In Ref.~\cite{Anselmino:2005ea} predictions for Sivers
SSAs in Drell-Yan processes at RHIC and PAX, based on a first extraction of
the Sivers functions from SIDIS data, with no sea contributions, were given.
A similar procedure, for RHIC experiments, was followed in
Ref.~\cite{Vogelsang:2005cs}. In Ref.~\cite{Collins:2005rq} estimates for SSAs
at RHIC, PAX (in fixed target mode) and COMPASS were given based on a simple
expression of the (opposite) $u$ and $d$ Sivers functions, extracted from a
two-parameter fit of the SIDIS HERMES data, and combined with some models for
the antiquark Sivers distributions. Here, we use the
recent~\cite{Anselmino:2008sg}, much more structured, valence and sea, Sivers
functions obtained from a fit of SIDIS asymmetries for pion and kaon
production, measured by HERMES and COMPASS collaborations. We give estimates
for the full set of experiments which either have been proposed or are being
discussed in many laboratories worldwide.

\section{\label{Mod} Formalism for the Sivers effect in Drell-Yan processes}

We consider the leading order (LO) parton model cross section for Drell-Yan
processes, $\Aup\,B \to \ell^+\ell^-X$ ($\Aup = \pup; \> B = p,\bar p,\pi)$,
in the hadronic c.m.~frame, in which one only observes the four-momentum of
the final $\ell^+\ell^-$ pair, or related quantities:
\be
q = (q_0, \bfq_T, q_L) \quad\quad q^2 = M^2 \quad\quad
y = \frac 12 \ln \frac{q_0 + q_L}{q_0 - q_L} \quad \quad
x_F = \frac{2\,q_L}{\sqrt s}\quad\quad s=(p_A+p_B)^2 \>\cdot
\label{var}
\ee
In the kinematical region
\be
q_T^2 \ll M^2 \quad\quad\quad k_{\perp} \simeq q_T \>, \label{kinr}
\ee
where $q_T = |\bfq_T|$ and $k_\perp = |\bfk_\perp|$ is the magnitude of the
parton intrinsic motion, at $O(k_\perp/M)$ the Sivers SSA is simply given by \cite{Anselmino:2002pd}
\bea
A_N &=& \frac{d\sigma^{A^\uparrow B \to \ell^+ \ell^- X}
          - d\sigma^{A^\downarrow B \to \ell^+ \ell^- X}}
           {d\sigma^{A^\uparrow B \to \ell^+ \ell^- X}
          + d\sigma^{A^\downarrow B \to \ell^+ \ell^- X}}
\equiv \frac{d\sigma^\uparrow - d\sigma^\downarrow}
           {d\sigma^\uparrow + d\sigma^\downarrow} \label{asy} \\
&=& \frac
{\sum_q \int d^2\bfk_{\perp 1} \, d^2\bfk_{\perp 2} \>
\delta^2(\bfk_{\perp 1} + \bfk_{\perp 2} - \bfq_T) \>
\Delta^N\!f_{q/\Aup}(x_1, \bfk_{\perp 1}) \>
 f_{\bar q/B}(x_{2}, k_{\perp 2})
\> \hat{\sigma}_0^{q\bar q}}
{2 \sum_q \int d^2\bfk_{\perp 1} \, d^2\bfk_{\perp 2} \>
\delta^2(\bfk_{\perp 1} + \bfk_{\perp 2} - \bfq_T) \>
 f_{q/A}(x_1, k_{\perp 1}) \>
  f_{\bar q/B}(x_{2}, k_{\perp 2})
\> \hat{\sigma}_0^{q\bar q}} \label{ann} \>,
\eea
with
\be
\hat{\sigma}_0^{q\bar q} = e_q^2 \, \frac{4\pi\alpha^2}{9 M^2}
\quad\quad\quad\quad
x_{1,2} = \frac{M}{\sqrt s} \, e^{\pm y}  =
\frac{\pm x_F + \sqrt{x_F^2 + 4 \, M^2/s}}{2} \>, \label{xqxqb}
\ee
where the subscripts $1$ and $2$ label the quark and antiquark participating
in the elementary LO scattering process, $q\bar q \to \ell^+ \ell ^-$. Notice
that, with the definition of $x_F$ adopted in Eq.~(\ref{var}), one has
\be
x_F = x_1 - x_2 \quad\quad\quad\quad\quad |x_F| \leq 1 - \frac{M^2}{s} \> \cdot
\ee

The sum in Eq.~(\ref{ann}) runs over all quarks and antiquarks
($q = u, \bar u, d,\bar d, s, \bar s$) and $d\sigma$ stands for
\be
\frac{d^4\sigma}{dy \, dM^2 \, d^2\bfq_T} =
\frac{1}{s} \, \frac{d^4\sigma}{d x_1 \, d x_{2} \, d^2\bfq_T} =
(x_1 + x_{2}) \, \frac{d^4\sigma}{d x_F \, dM^2 \, d^2\bfq_T} =
\frac{1}{2} \, \frac{d^4\sigma}{d^4q} \cdot \label{x-sect}
\ee

Notice that in obtaining Eq.~(\ref{ann}) one uses (see Eq.~(\ref{sivnoi})):
\bea
\!\!\!\!\!\!\!\!\!\!
\hat f_{q/\Aup}(x, \bfk_{\perp}) + \hat f_{q/\Adown}(x, \bfk_{\perp})
\!\! &=& \!\! 2\, f_{q/A}(x, k_{\perp}) \\
\!\!\!\!\!\!\!\!\!\!
\hat f_{q/\Aup}(x, \bfk_{\perp}) - \hat f_{q/\Adown}(x, \bfk_{\perp})
\!\! &=& \!\! \Delta^N\!f_{q/\Aup}(x, k_{\perp}) \> \bfS \cdot
(\hat{\bfP} \times \hat{\bfk}_{\perp}) \equiv
\Delta^N\!f_{q/\Aup}(x, \bfk_{\perp}) \,. \label{f-f}
\eea
The SSA (\ref{ann}) depends directly on the Sivers functions
$\Delta^N\!f_{q/\Aup}$.

We use here the same factorized expression of the cross sections which
holds in the collinear configuration, generalizing it to the case of
unintegrated, or transverse momentum dependent (TMD), partonic
distributions~\cite{D'Alesio:2004up,D'Alesio:2007jt}. A most general
treatment of unpolarized and polarized Drell-Yan processes in such a
scheme has very recently appeared~\cite{Arnold:2008kf}.
Factorization for SIDIS and Drell-Yan processes in the kinematical
regime we are considering here, Eq.~(\ref{kinr}), has been proven in QCD
\cite{Collins:1984kg,Ji:2004wu,Ji:2004xq}, resulting in the same parton
model TMD-factorization adopted here and in Ref.~\cite{Arnold:2008kf}, with
the addition of an extra soft factor $S$, which takes into account transverse
motion originating from soft gluon emission (see, for example, Eqs.~(40)
and~(41) of Ref.~\cite{Ji:2006vf}). Such a factor gives an (unknown)
additional contribution, both to the numerator and denominator of
Eq.~(\ref{ann}), of order $\alpha_s$, and is neglected here. The TMD
partonic distributions we use are obtained by fitting experimental data
and take into account all sources of intrinsic motion.

\subsection{About sign and azimuthal angle conventions}

As the issue of the sign of the Sivers asymmetry in Drell-Yan processes
is so important, let us discuss in details the choices adopted here and their
motivation. We define our kinematical configuration with hadron $\Aup$ moving
along the positive $z$-axis, and hadron $B$ opposite to it, in the $A-B$
center of mass frame. We choose the ``up" $(\uparrow)$ polarization direction
as the positive $y$-axis ($\phi_S = \pi/2$). The transverse momenta have
azimuthal angles
\be
\bfq_T = q_T(\cos\phi_\gamma, \, \sin\phi_\gamma,\, 0) \quad\quad\quad
\bfk_{\perp i} = k_{\perp i}(\sin\varphi_i, \, \cos\varphi_i, \, 0)
\quad\quad (i = 1,2) \>, \label{phi-dep}
\ee
so that the mixed product $\bfS \cdot (\hat{\bfP} \times \hat{\bfk}_{\perp 1})$
gives an azimuthal dependence $\sin(\phi_S - \varphi_1) = \cos\varphi_1$ which,
upon integration on $\bfk_{\perp 1}$, yields a $\sin(\phi_S - \phi_{\gamma}) =
\cos\phi_\gamma$ dependence of the Sivers asymmetry (see Eq.~(\ref{num}) below).

Notice that, contrary to the usual study of angular dependences of Drell-Yan
processes which is mainly performed in the so-called  Collins-Soper reference
frame \cite{Collins:1977iv}, we work in the hadronic c.m.~frame. At least for
the purpose of studying the Sivers asymmetry, this frame is much more convenient
and directly related to experimental measurements.

In order to collect data at all azimuthal angles, following what is usually
done in semi-inclusive deep inelastic scattering processes, both experimental
results and theoretical estimates can be discussed for the azimuthal moments of
the asymmetry. We follow Refs.~\cite{Efremov:2004tp,Collins:2005rq}
and~\cite{Vogelsang:2005cs} and choose as a weight the
$\sin(\phi_{\gamma} - \phi_S)$ phase. We then have:
\bea
&& A_N^{\sin(\phi_{\gamma} - \phi_S)} \equiv
\frac{\int_0^{2\pi} d\phi_{\gamma} \>
[d\sigma^{\uparrow} - d\sigma^{\downarrow}] \>
\sin(\phi_{\gamma}-\phi_S)}
{\frac{1}{2}\int_{0}^{2\pi} d\phi_{\gamma} \>
[d\sigma^{\uparrow} + d\sigma^{\downarrow}]} \label{ANW} \\
&=& \frac{\int d\phi_{\gamma} \> \left[
\sum_q \int d^2\bfk_{\perp 1} \, d^2\bfk_{\perp 2} \>
\delta^2(\bfk_{\perp 1} + \bfk_{\perp 2} - \bfq_T) \>
\Delta^N\!f_{q/\Aup}(x_1, \bfk_{\perp 1}) \>
 f_{\bar q/B}(x_{2}, k_{\perp 2})\>
\hat{\sigma}_0^{q\bar q}\right] \> \sin(\phi_{\gamma}-\phi_S)}
{\int d\phi_{\gamma} \> \left[
\sum_q \int d^2\bfk_{\perp 1} \, d^2\bfk_{\perp 2} \>
\delta^2(\bfk_{\perp 1} + \bfk_{\perp 2} - \bfq_T) \>
 f_{q/A}(x_1, k_{\perp 1}) \>
 f_{\bar q/B}(x_{2}, k_{\perp 2})\>\hat{\sigma}_0^{q\bar q}
\right]} \> \cdot \nonumber
\eea
Notice that such a choice, combined with the $\sin(\phi_S - \phi_{\gamma})$
dependence associated with the Sivers function [see comments following
Eq.~(\ref{phi-dep})] implies an overall $[-\sin^2(\phi_{\gamma} - \phi_S)]$
factor in the numerator of Eq.~(\ref{ANW}).

The above asymmetry is, in general, a function of $x_F$ (or $y), \, M$ and $q_T$. In
the sequel we shall study it as a function of one variable only, either $x_F$
or $M$; we will always integrate both the numerator and denominator of
Eq.~(\ref{ANW}) over $q_T$ --- covering the range in which the factorized
approach with unintegrated distribution functions is supposed to hold
(as detailed in the captions of Figs.~\ref{Compass-AN}-\ref{PRO-AN}) --- and on one of the remaining
variables according to the kinematical ranges of the corresponding experiments.

The case in which the polarized hadron $A^\uparrow$ moves along $-\hat{\bfz}$,
that is the process $B \, \Aup \to \ell^+\ell^-X$, deserves a special comment.
In such a case --- {\it keeping the same definition of $\uparrow$ polarization
and the same reference frame} --- the Sivers mixed product has an opposite sign
(due to the opposite $\Aup$ momentum) and yields a Sivers asymmetry proportional
to $\sin(\phi_{\gamma}-\phi_S)$. In this case the overall factor in the numerator
of Eq.~(\ref{ANW}) is $[+\sin^2(\phi_{\gamma} - \phi_S)]$. This agrees with
what is done in SIDIS processes, $\ell \, \pup \to \ell \, h \, X$, in the
$\gamma^* - \pup$ c.m.~frame~\cite{Bacchetta:2004jz}. To summarize, we shall give
estimates for the quantities:
\be
A_N^{\sin(\phi_{\gamma} - \phi_S)}(\Aup \, B \to \gamma^* X;
\> x_F, M, q_T) =
- A_N^{\sin(\phi_{\gamma} - \phi_S)}(B \, \Aup \to \gamma^* X;
\> -x_F, M, q_T) \>. \label{ab-ba}
\ee
The equality holds due to rotational invariance.

\section{Estimates for forthcoming experiments}\label{est}

In order to give estimates for the Sivers asymmetries in Drell-Yan processes ---
and test the crucially important sign change when going from SIDIS to DY ---
we only need to insert the Sivers functions extracted from the analysis of SIDIS
data into Eq.~(\ref{ANW}). We use the results obtained in
Ref.~\cite{Anselmino:2008sg}, which adopted a Gaussian factorized form both
for the unpolarized distribution functions:
\be
f_{q/p}(x,k_\perp) = f_q(x) \, \frac{1}{\pi \langle\kt^2\rangle} \,
e^{-{\kt^2}/{\langle\kt^2\rangle}} \quad\quad
\langle\kt^2\rangle   = 0.25  \;{\rm GeV}^2 \>,
\label{partond}
\ee
and for the Sivers distributions:
\bea
\Delta^N \! f_ {q/\pup}(x,\kt) &=& 2 \, {\cal N}_q(x) \, h(\kt) \,
f_ {q/p} (x,\kt) \label{sivfac} \\
&\equiv& \Delta^N \! f_ {q/\pup}(x) \, h(\kt) \,
\frac{1}{\pi \langle\kt^2\rangle} \, e^{-{\kt^2}/{\langle\kt^2\rangle}}
\nonumber \; ,
\eea
where
\bea
&&{\cal N}_q(x) =  N_q \, x^{\alpha_q}(1-x)^{\beta_q} \,
\frac{(\alpha_q+\beta_q)^{(\alpha_q+\beta_q)}}
{\alpha_q^{\alpha_q} \beta_q^{\beta_q}}
\label{siversx} \\
&&h(\kt) = \sqrt{2e}\,\frac{k_\perp}{M_{1}}\,e^{-{k_\perp^2}/{M_{1}^2}}\> \cdot
\label{siverskt}
\eea
The values of the 11 best fit parameters $N_q \,(q=u,d,s,\bar u, \bar d, \bar s)$,
$\alpha_q \, (q=u, d, sea)$, $\beta$ (same for all $q$) and $M_1$ can be found
in Table I of Ref.~\cite{Anselmino:2008sg}, where their uncertainty is also
explained in details.

Notice that the above factorized expressions allow, at {$\cal O$}($k_\perp/M$), an
analytical integration of the numerator and denominator of Eq.~(\ref{ANW}),
resulting in
\bea
A_N^{\sin(\phi_{\gamma} - \phi_S)}(x_F, M, q_T) =
\frac{\int d\phi_\gamma \> [N(x_F, M, q_T, \phi_\gamma)] \>
\sin(\phi_{\gamma} - \phi_S)}
{\int d\phi_\gamma \> [D(x_F, M, q_T)]} \label{an-gauss}
\eea
with (see Eq.~(\ref{x-sect})):
\bea
N(x_F, M, q_T, \phi_\gamma)
&\equiv& \frac{d^4\sigma^{\uparrow}}{dx_F \, dM^2 \, d^2\bfq_T}
- \frac{d^4\sigma^{\downarrow}}{dx_F \, dM^2 \, d^2\bfq_T}   \nonumber \\
&& \hspace*{-1.8cm}=\frac{4 \, \pi \, \alpha^2}{9 \, M^2 \, s}\> \sum_q \>
\frac{e_q^2}{x_1+x_2} \,
\Delta^N f_{q/A^{\uparrow}}(x_1) \> f_{\bar{q}/B} (x_{2})\> \sqrt{2e} \>
\frac{q_T }{M_1} \, \frac{\langle k_{S}^2\rangle^2
\> \exp\left[\,-q_T^2/\left(\,\langle k_{S}^2\rangle + \langle k_{\perp 2}^2\rangle\,\right)\,\right]}
{\pi\left[ \langle k_{S}^2\rangle + \langle k_{\perp 2}^2\rangle
\right]^2\langle k_{\perp 2}^2\rangle} \> \sin(\phi_S- \phi_{\gamma})
\nonumber \\ && \;\label{num}
\eea
and
\bea
D(x_F, M, q_T)
&\equiv& \frac{1}{2} \left[ \frac{d^4\sigma^{\uparrow}}{dx_F \, dM^2 \,
d^2\bfq_T}
+ \frac{d^4\sigma^{\downarrow}}{dx_F \, dM^2 \, d^2\bfq_T} \right] =
\frac{d^4\sigma^{unp}}{dx_F \, dM^2 \, d^2\bfq_T}  \nonumber \\
&=& \frac{4 \, \pi \, \alpha^2}{9 \, M^2 \, s} \> \sum_q \>
\frac{e_q^2}{x_1+x_2}
f_{q/A}(x_1) \> f_{\bar{q}/B}(x_{2}) \>
\frac{\exp\left[\,-q_T^2/\left(\,\langle k_{\perp 1}^2\rangle + \langle k_{\perp 2}^2\rangle\,\right)\,\right]}
{\pi\left[\langle k_{\perp 1}^2\rangle +
\langle k_{\perp 2}^2\rangle\right]} \> \cdot
\eea
Notice that we have defined
\be
\frac{1}{\langle k_{S}^2\rangle} =
\frac{1}{M_1^2}+\frac{1}{\langle k_{\perp 1}^2\rangle}
\ee
and that $x_1, x_{2}$ have the values given in Eq.~(\ref{xqxqb}). Although in
the above equations we have allowed for the possibility of having different
values of $\langle k_{\perp 1,2}^2 \rangle$ for hadrons $A$ and $B$, in our
numerical evaluations we shall take them equal, as given by
Eq.~(\ref{partond}), even when one of the hadrons is a pion.

\begin{figure}[t]
\includegraphics[width=0.72 \textwidth,bb= 10 40 540 740,angle=0]{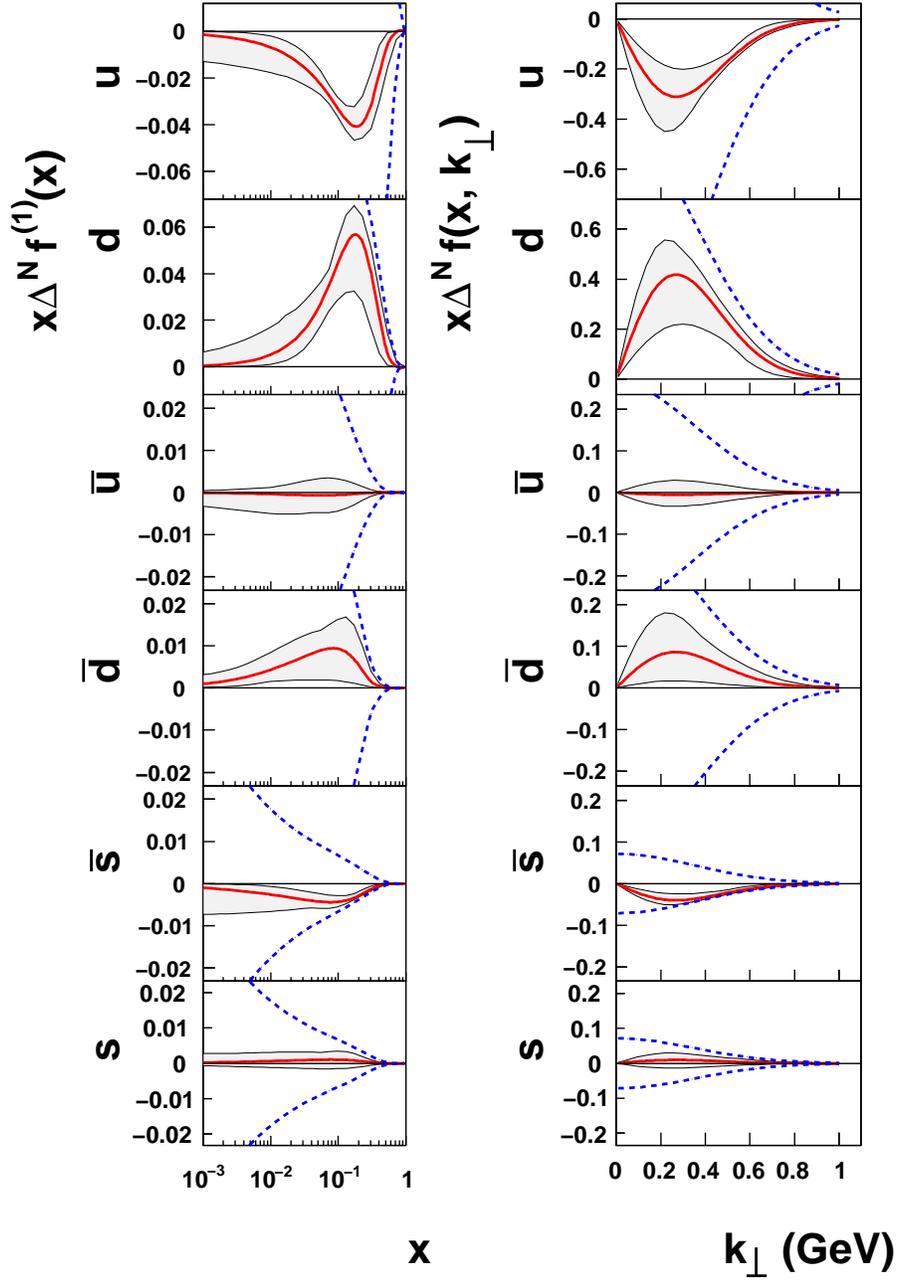}
\caption{\label{fig:sivers} The Sivers distribution functions for
$u$, $d$ and $s$ flavours as determined by our simultaneous fit of
HERMES and COMPASS data in Ref.~\cite{Anselmino:2008sg}. The sign
was reversed according to the prediction of
Refs.~\cite{Brodsky:2002rv,Collins:2002kn}. On the left panel, the
first moment, $x\,\Delta^N \! f^{(1)}(x)$, is shown as a function of
$x$ at $Q^2=2.4$~GeV$^2$ for each flavour. On the right panel, the
Sivers distribution, $x\,\Delta^N \! f(x,\kt)$, is shown as a
function of $\kt$ at a fixed value of $x = 0.1$ for each flavour. In
each plot, the highest and lowest dashed lines show the positivity
limits $|\Delta^N \! f| = 2f$. }
\end{figure}

\begin{center}
\begin{figure}
\hspace*{-1.3cm}
\includegraphics[angle=-90,width=0.45\textwidth]
{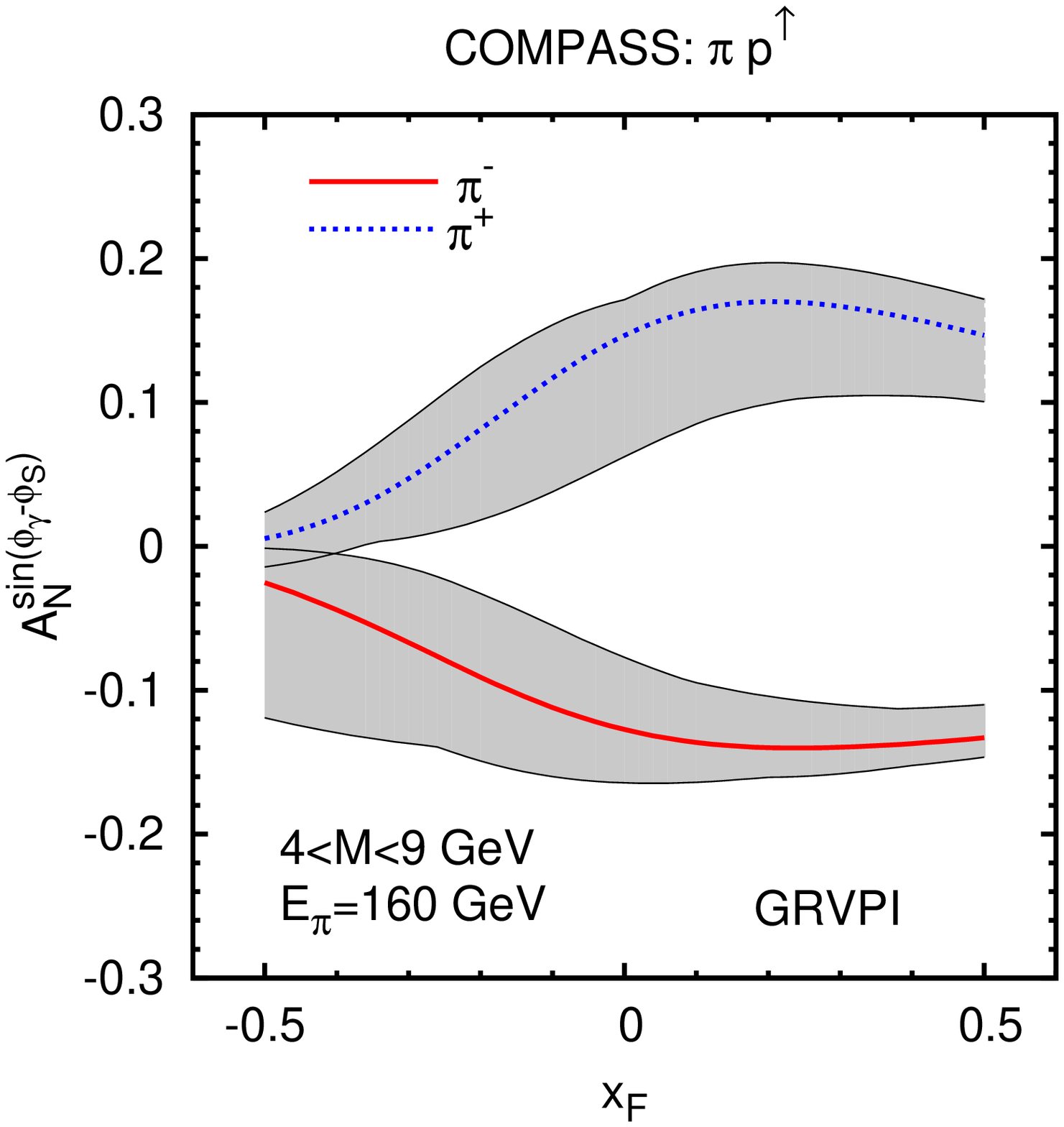}\hspace*{-2cm}
\includegraphics[angle=-90,width=0.433\textwidth]
{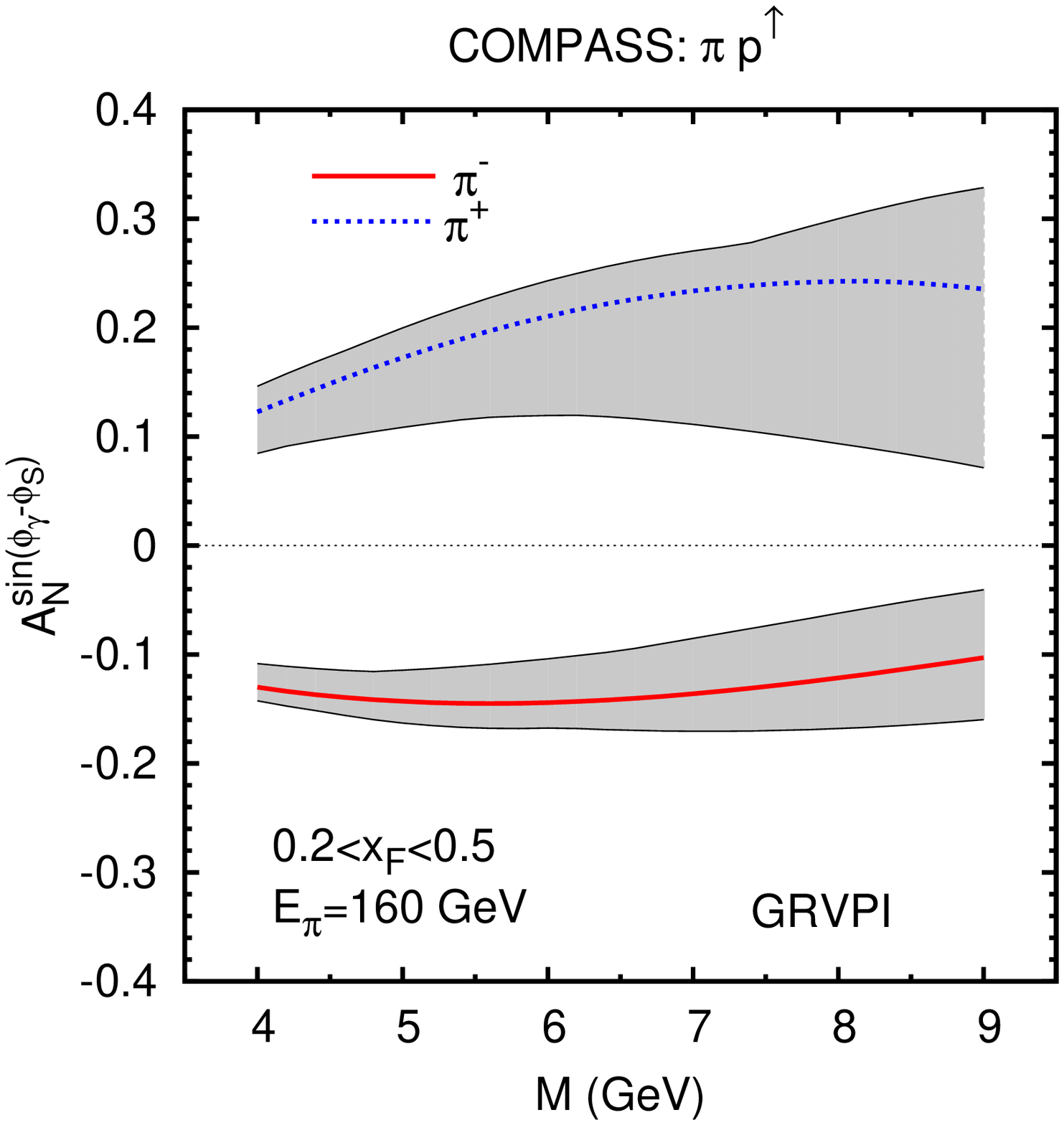}\hspace*{-2cm}
\includegraphics[angle=-90,width=0.45\textwidth]
{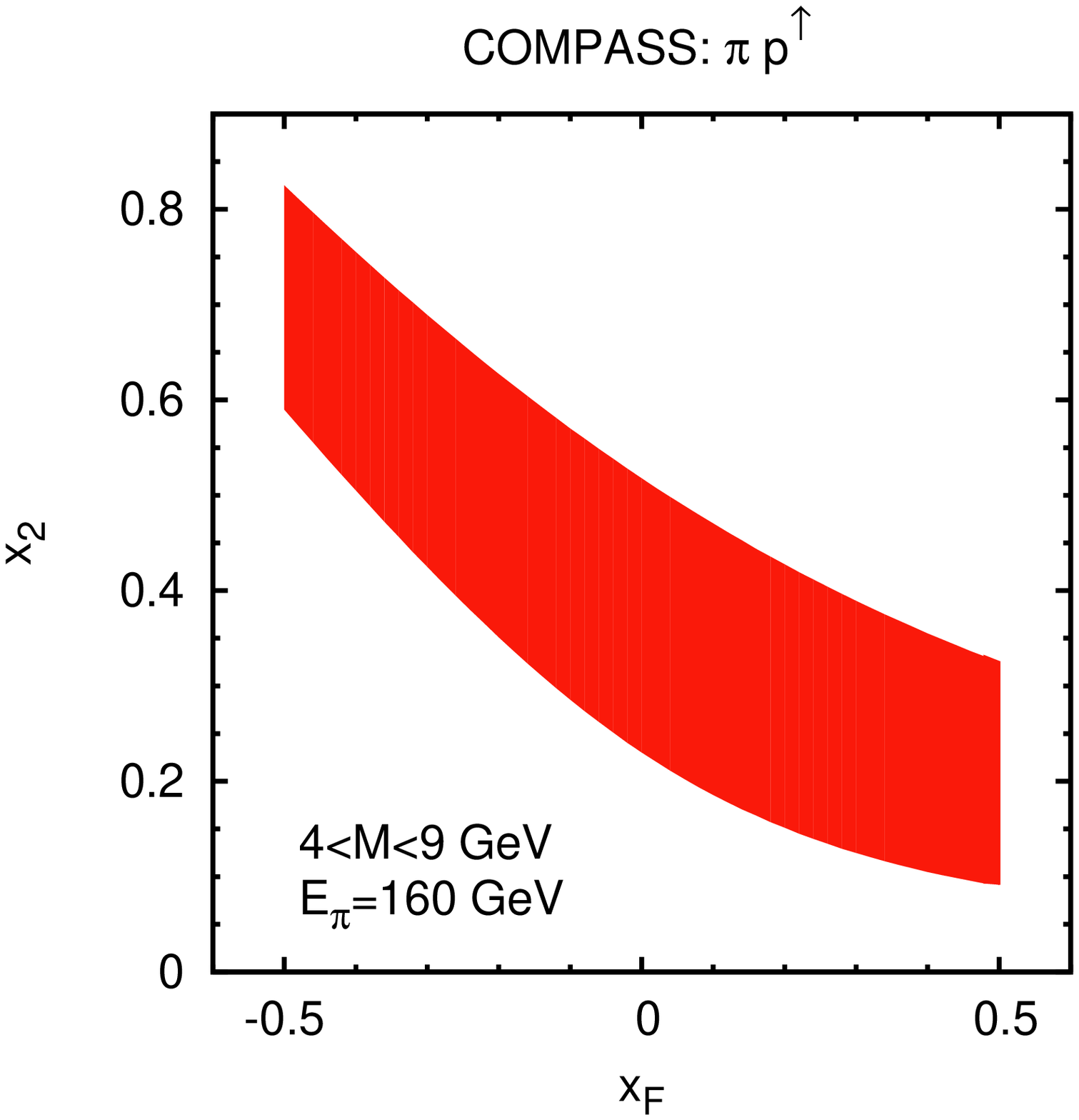}\hspace*{-2cm} \caption{The single spin
asymmetries $A_N^{\sin(\phi_{\gamma}-\phi_S)}$ for the Drell-Yan
process $\pi^\pm p ^\uparrow \to \mu^{+}\mu^{-}\,X$ at COMPASS, as a
function of $x_F = x_1 - x_{2}$ (left panel) and as a function of
$M$ (central panel). The integration ranges are $(0 \leq q_T \leq 1)$~GeV,
$(4 \leq M \leq 9)$~GeV and $0.2 \leq x_F \leq 0.5$. The results
are given for a pion beam energy of 160~GeV, corresponding to
$\sqrt{s} = 17.4$~GeV. The right panel shows the allowed region of
$x_2$ values as a function of $x_F$.}\label{Compass-AN}
\end{figure}
\end{center}

\begin{center}
\begin{figure}
\hspace*{-1.3cm}
\includegraphics[angle=-90,width=0.45\textwidth]
{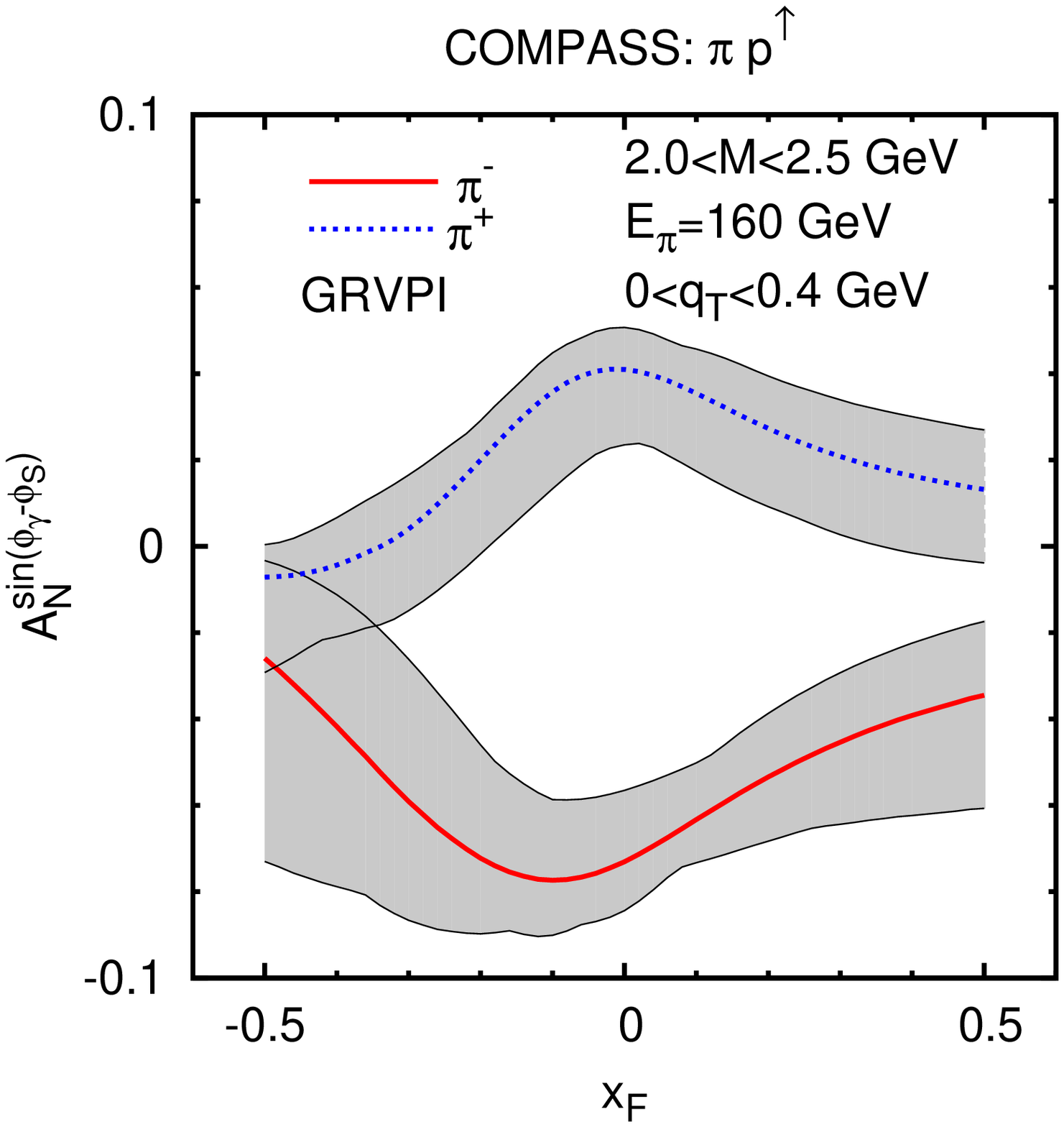}\hspace*{-2cm}
\includegraphics[angle=-90,width=0.45\textwidth]
{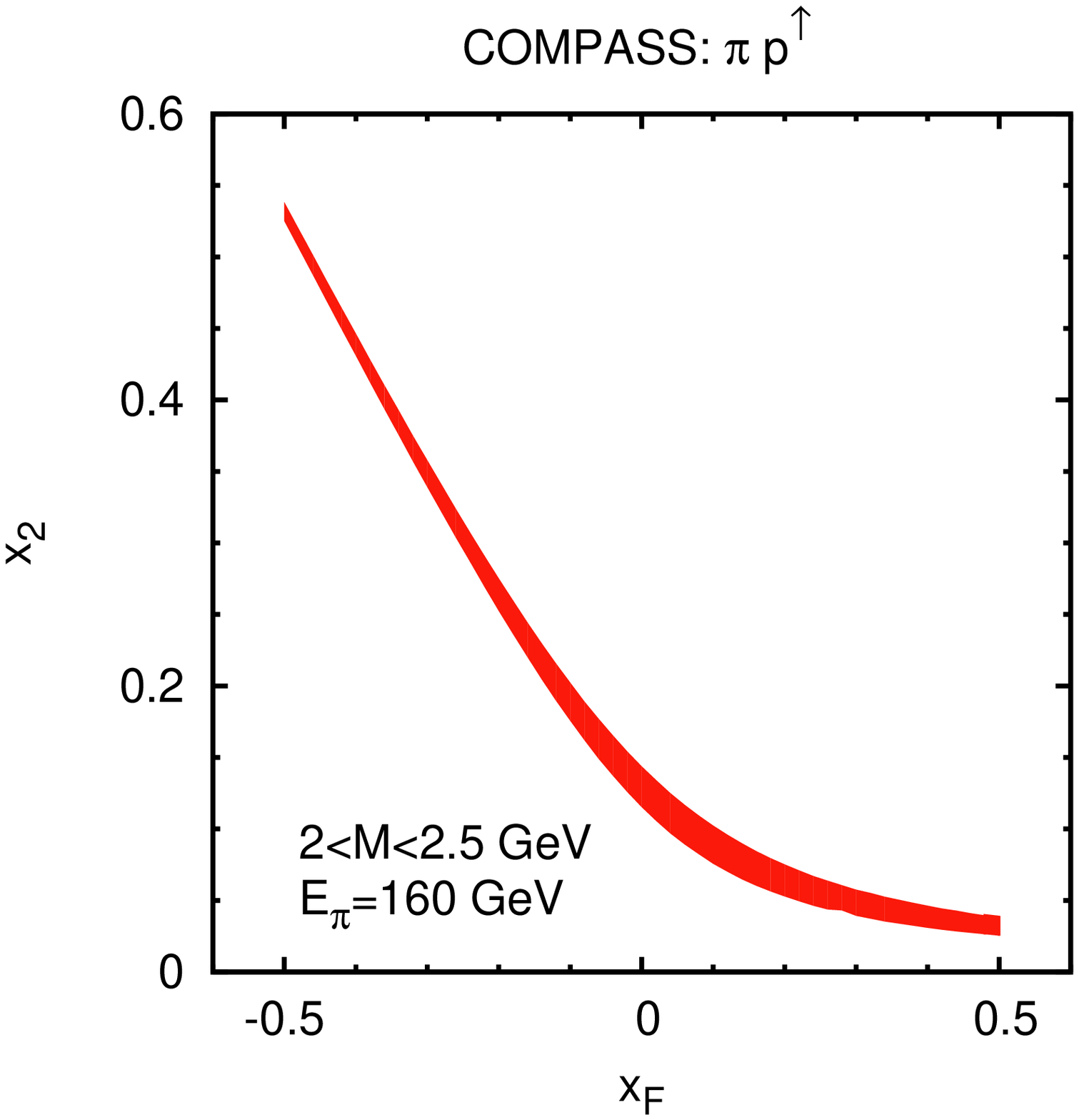} \caption{The single spin asymmetries $A_N^{\sin(\phi_{\gamma}-\phi_S)}$ for the Drell-Yan process
$\pi^\pm p ^\uparrow \to \mu^{+}\mu^{-}\,X$ at COMPASS, as a function of
$x_F = x_1 - x_{2}$ (left panel). The integration
ranges are $(0 \leq q_T \leq 0.4)$~GeV and $(2.0 \leq M \leq 2.5)$~GeV.
The results are given for a pion beam energy of 160~GeV, corresponding to
$\sqrt s = 17.4$~GeV. The right panel shows the allowed region of $x_2$
values as a function of $x_F$.}\label{Compass-AN-low}
\end{figure}
\end{center}

\begin{center}
\begin{figure}
\hspace*{-1.3cm}
\includegraphics[angle=-90,width=0.45\textwidth]
{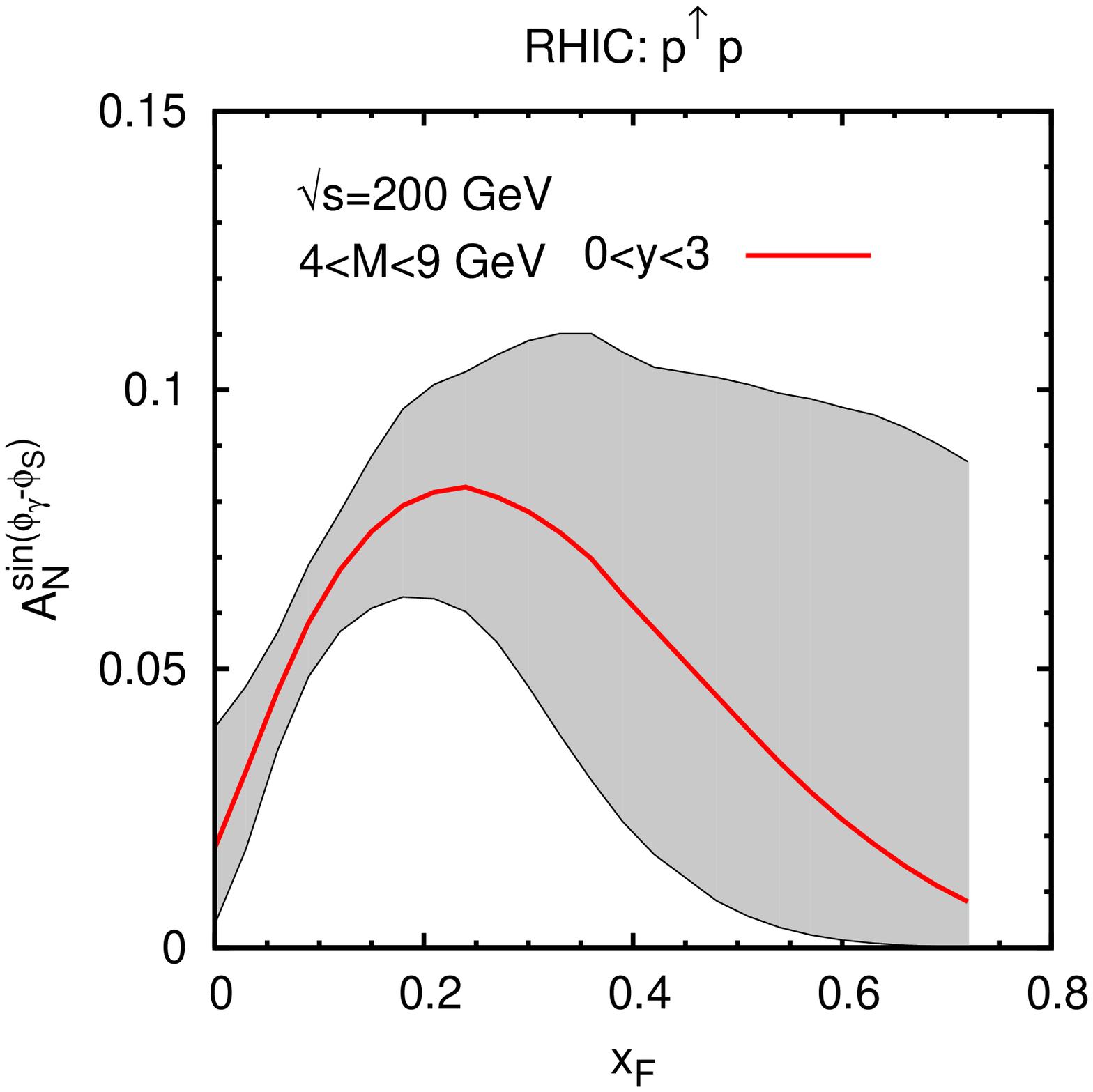}\hspace*{-2cm}
\includegraphics[angle=-90,width=0.433\textwidth]
{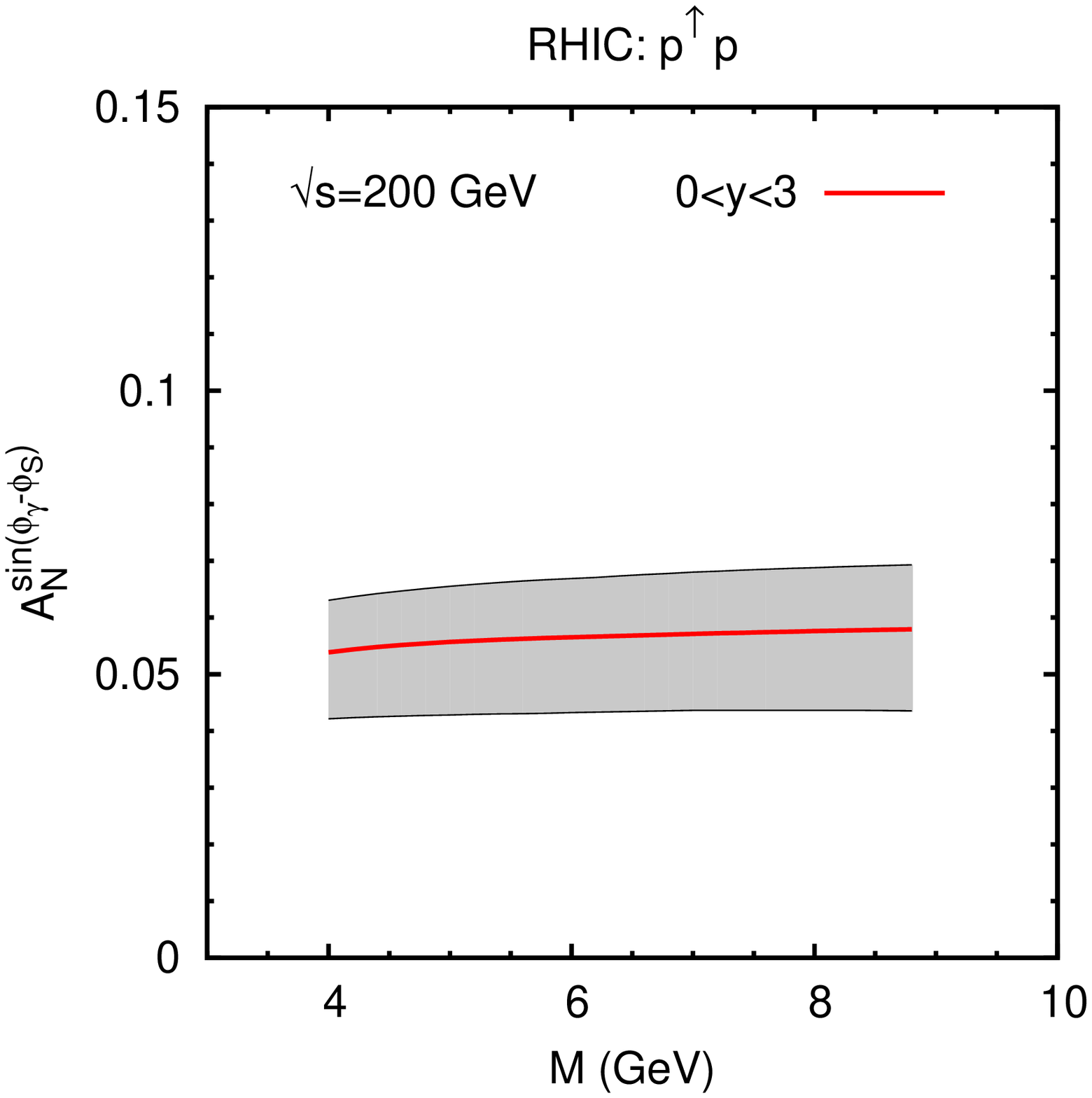}\hspace*{-2.cm}
\includegraphics[angle=-90,width=0.45\textwidth]
{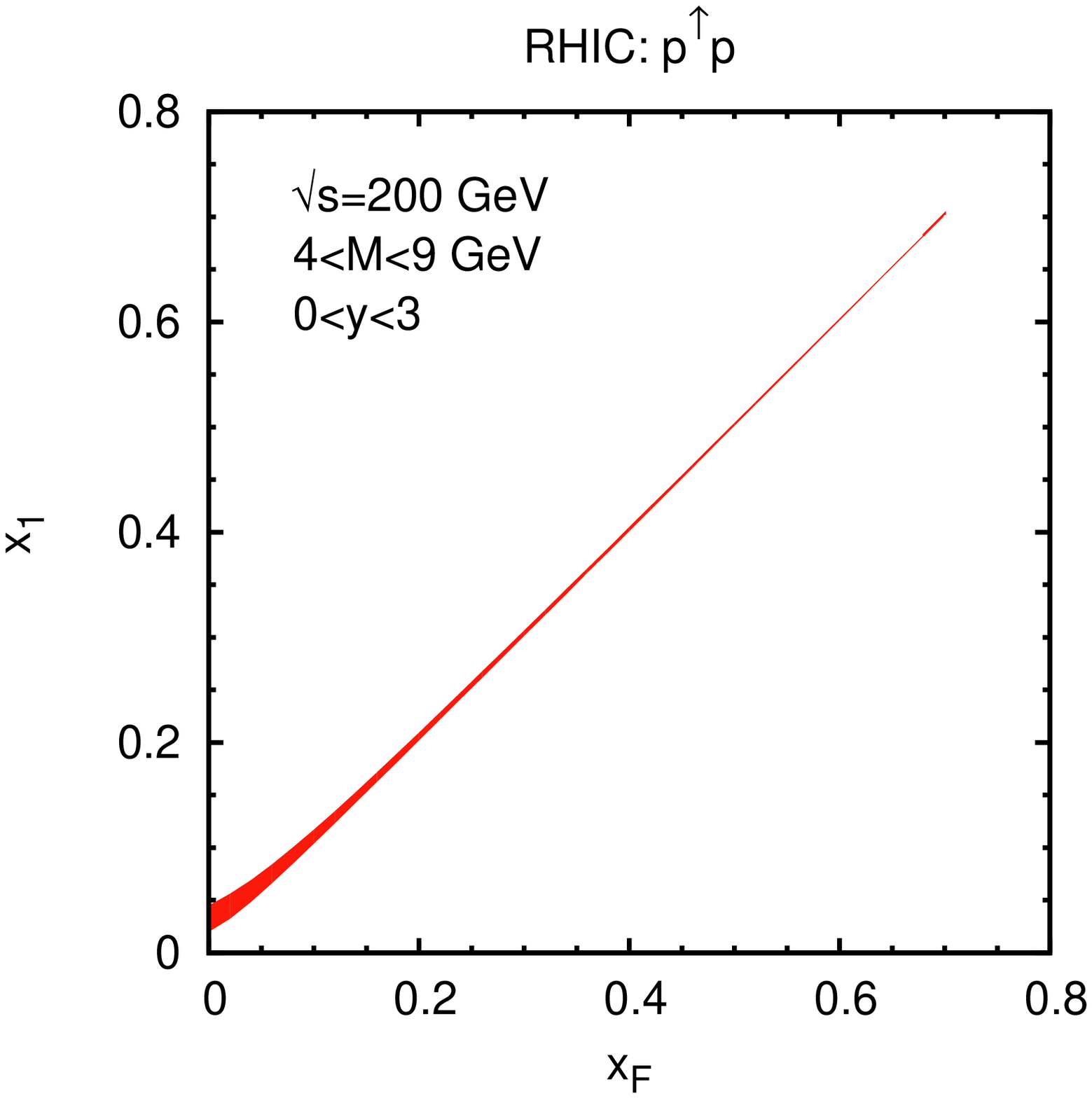}\hspace*{-2cm} \caption{The single spin
asymmetry $A_N^{\sin(\phi_{\gamma}-\phi_S)}$ for the Drell-Yan
process $p^{\uparrow} p\to \mu^{+}\mu{^-}\,X$ at RHIC, as a function
of $x_F$ (left panel) and $M$ (central panel). The integration
ranges are $(0 \leq q_T \leq 1)$~GeV and $(4 \leq M \leq 9)$~GeV, with the
further constraint $0 \leq y \leq 3$. The results are given at
$\sqrt s = 200$~GeV. The right panel shows the allowed region of
$x_1$ values as a function of $x_F$.} \label{RHIC-AN}
\end{figure}
\end{center}

\begin{center}
\begin{figure}
\hspace*{-1.3cm}
\includegraphics[angle=-90,width=0.45\textwidth]
{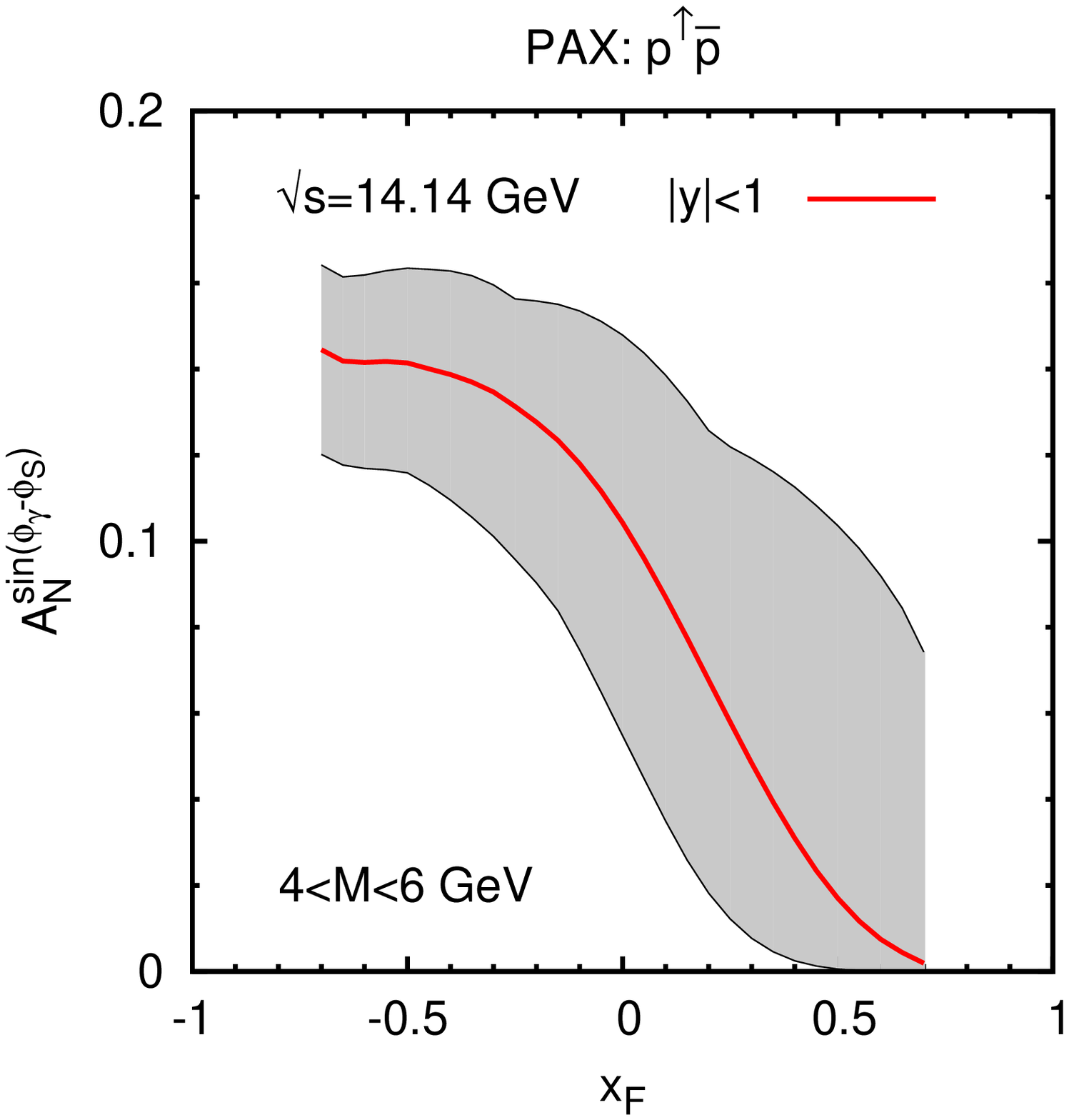}\hspace*{-2cm}
\includegraphics[angle=-90,width=0.435\textwidth]
{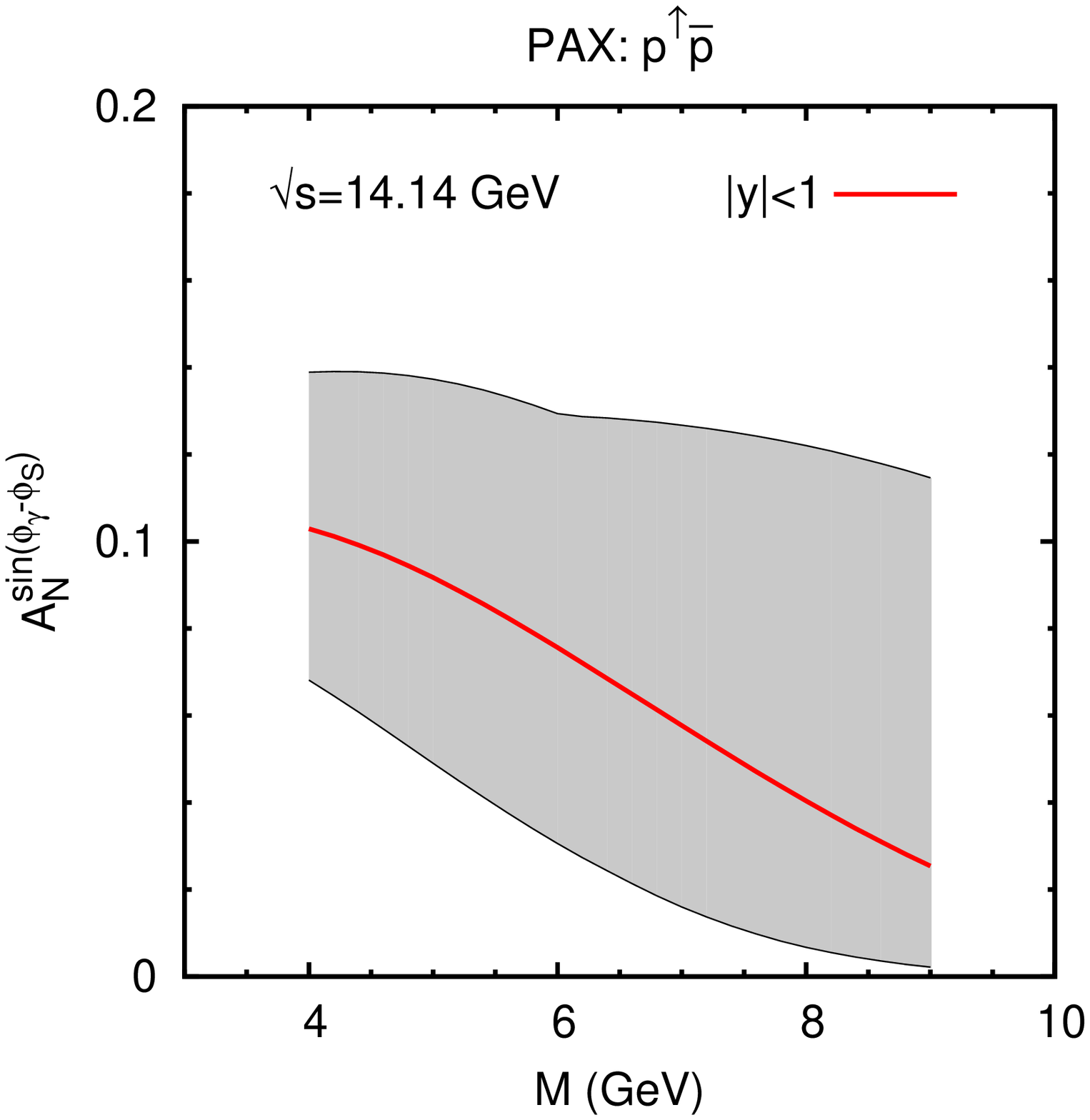}\hspace*{-2cm}
\includegraphics[angle=-90,width=0.45\textwidth]
{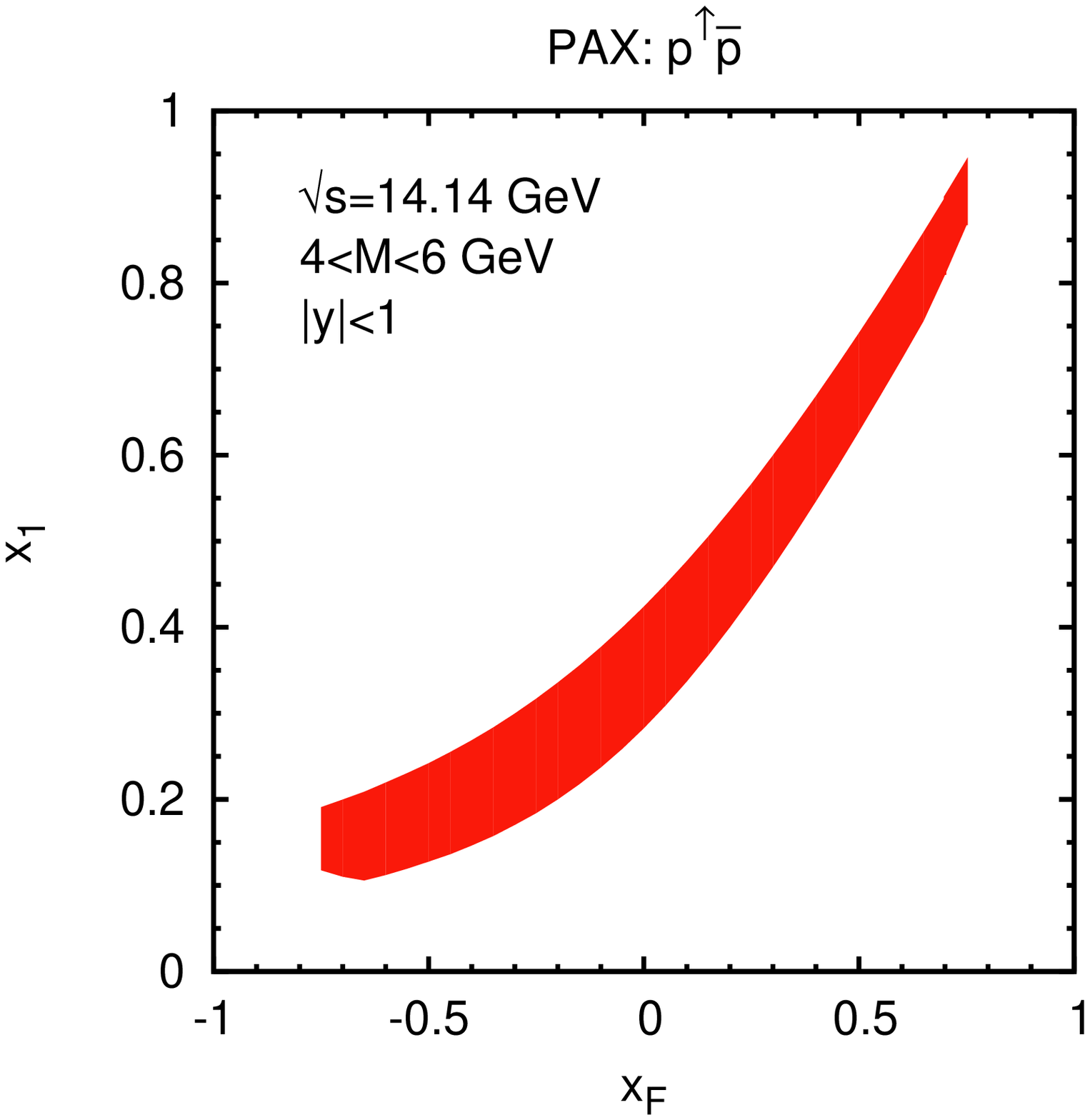} \caption{The single spin asymmetry
$A_N^{\sin(\phi_{\gamma}-\phi_S)}$ for the Drell-Yan process
$p^{\uparrow} \bar{p}\to \mu^{+}\mu^{-}\,X$ at PAX, as a function of
$x_F$ (left panel) and $M$ (central panel). The integration ranges are
$(0 \leq q_T \leq 1)$~GeV and $(4 \leq M \leq 6)$~GeV with the further
constraint $|y| \leq 1$. The results are given at $\sqrt s = 14.14$~GeV.
The right panel shows the allowed region of $x_1$ values as a
function of $x_F$.}\label{PAX-AN}
\end{figure}
\end{center}

\begin{center}
\begin{figure}
\includegraphics[angle=-90,width=0.45\textwidth]
{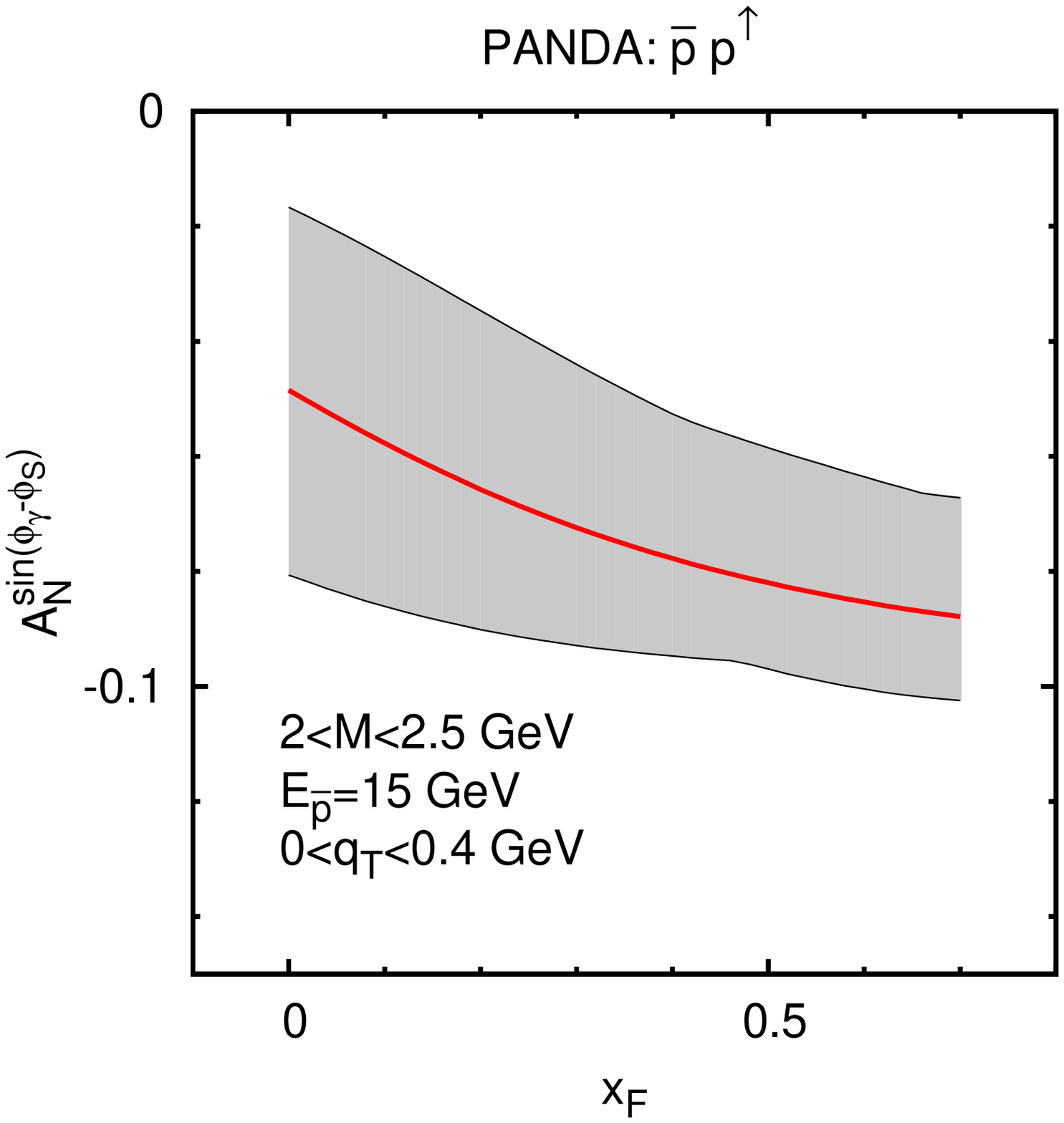}
\includegraphics[angle=-90,width=0.45\textwidth]
{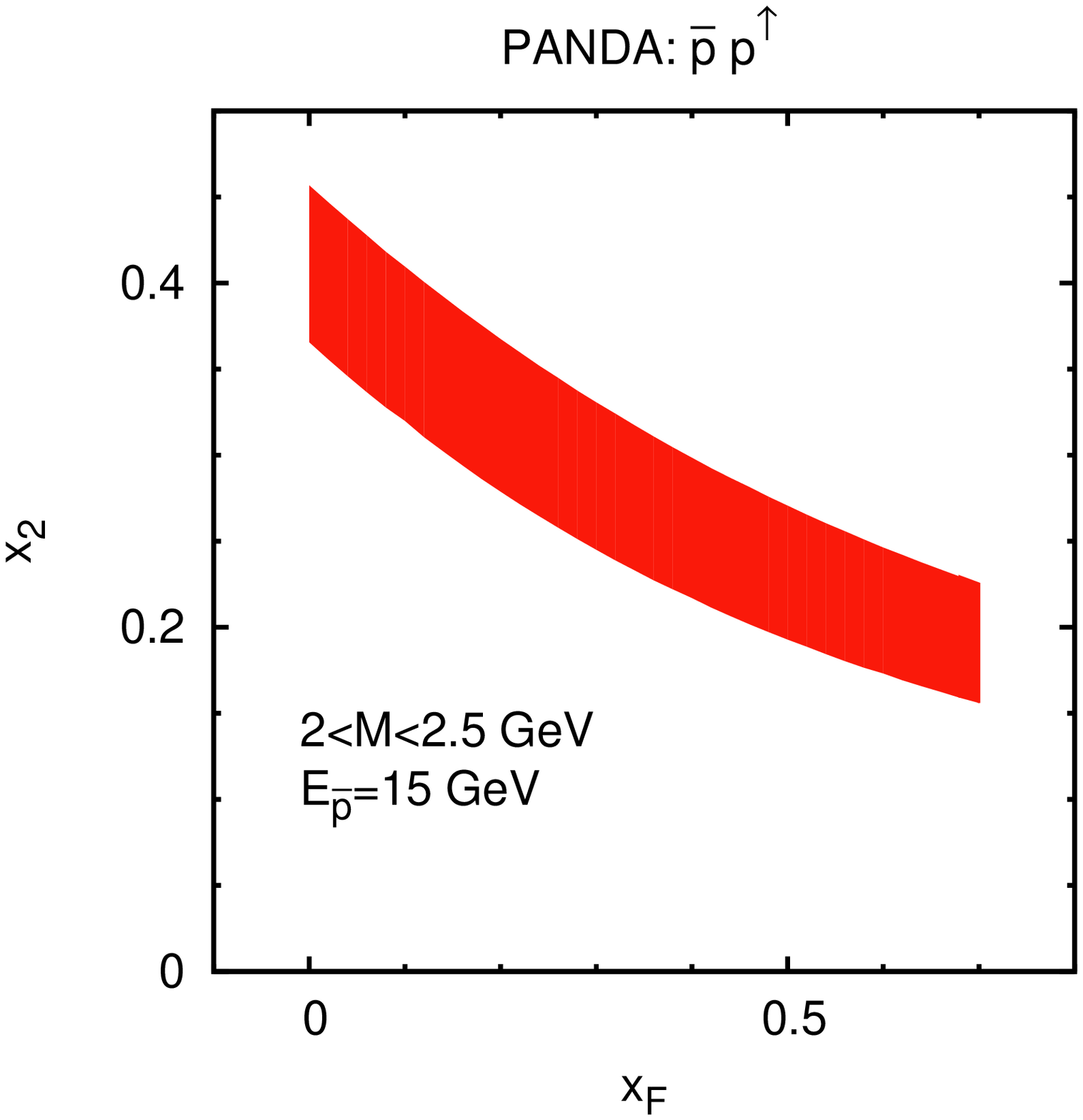} \caption{The single spin asymmetry
$A_N^{\sin(\phi_{\gamma}-\phi_S)}$ for the Drell-Yan process
$\bar{p} p^{\uparrow}\to \mu^{+}\mu^{-}\,X$ at PANDA, as a function
of $x_F$. The integration ranges are $(0 \leq q_T \leq 0.4)$~GeV and
$(2 \leq M \leq 2.5)$~GeV. The right panel shows the allowed region
of $x_2$ values as a function of $x_F$. The results are given at $E_{\bar p}
= 15$~GeV, corresponding to $\sqrt s= 5.47$~GeV.}\label{PANDA-AN}
\end{figure}
\end{center}

\begin{center}
\begin{figure}
\includegraphics[angle=-90,width=0.45\textwidth]{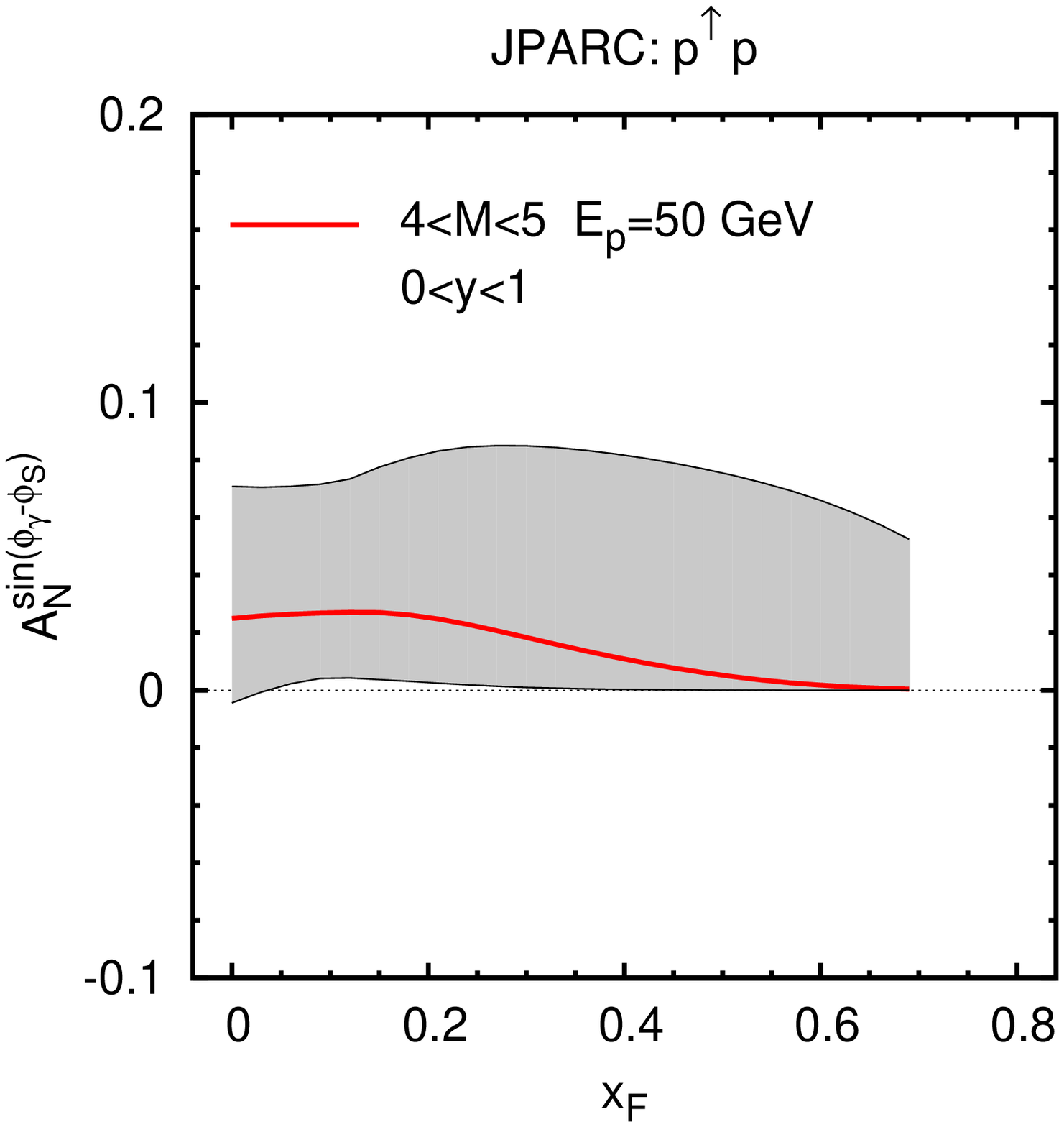}
\includegraphics[angle=-90,width=0.45\textwidth]{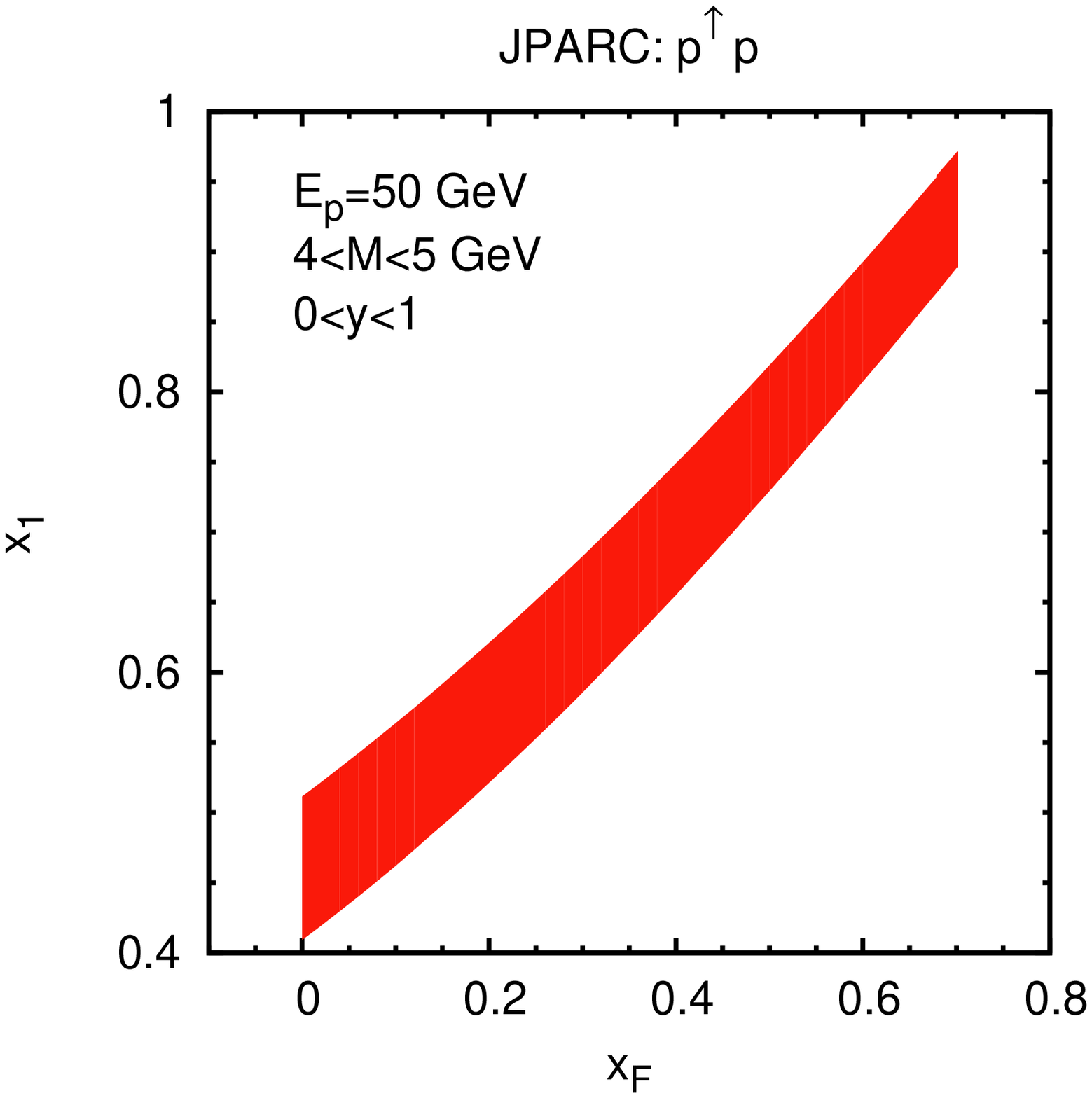}
\caption{The single spin asymmetry
$A_N^{\sin(\phi_{\gamma}-\phi_S)}$ for the Drell-Yan process
$p^\uparrow p\to \mu^{+}\mu^{-}\,X$ at J-PARC, as a function of
$x_F$. The integration ranges are $(0 \leq q_T \leq 1)$~GeV and
$(4 \leq M \leq 5)$~GeV.
The results are given at $E_p = 50$~GeV, corresponding to $\sqrt
s=9.78$~GeV. The right panel shows the allowed region of $x_1$
values as a function of $x_F$.}\label{JPARC-AN}
\end{figure}
\end{center}

\begin{center}
\begin{figure}
\hspace*{-2cm}
\includegraphics[angle=-90,width=0.45\textwidth]
{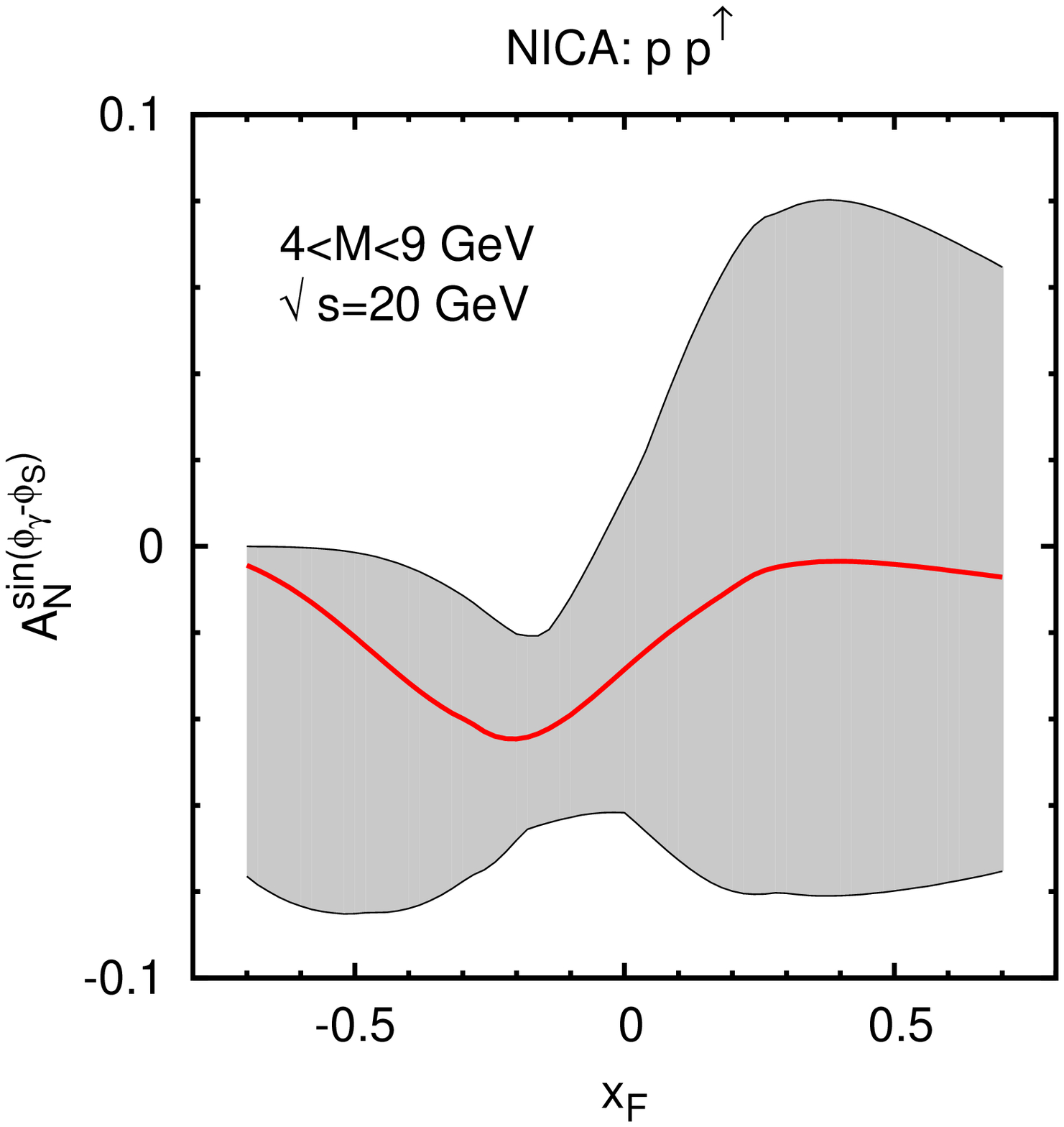}\hspace*{-2cm}
\includegraphics[angle=-90,width=0.45\textwidth]
{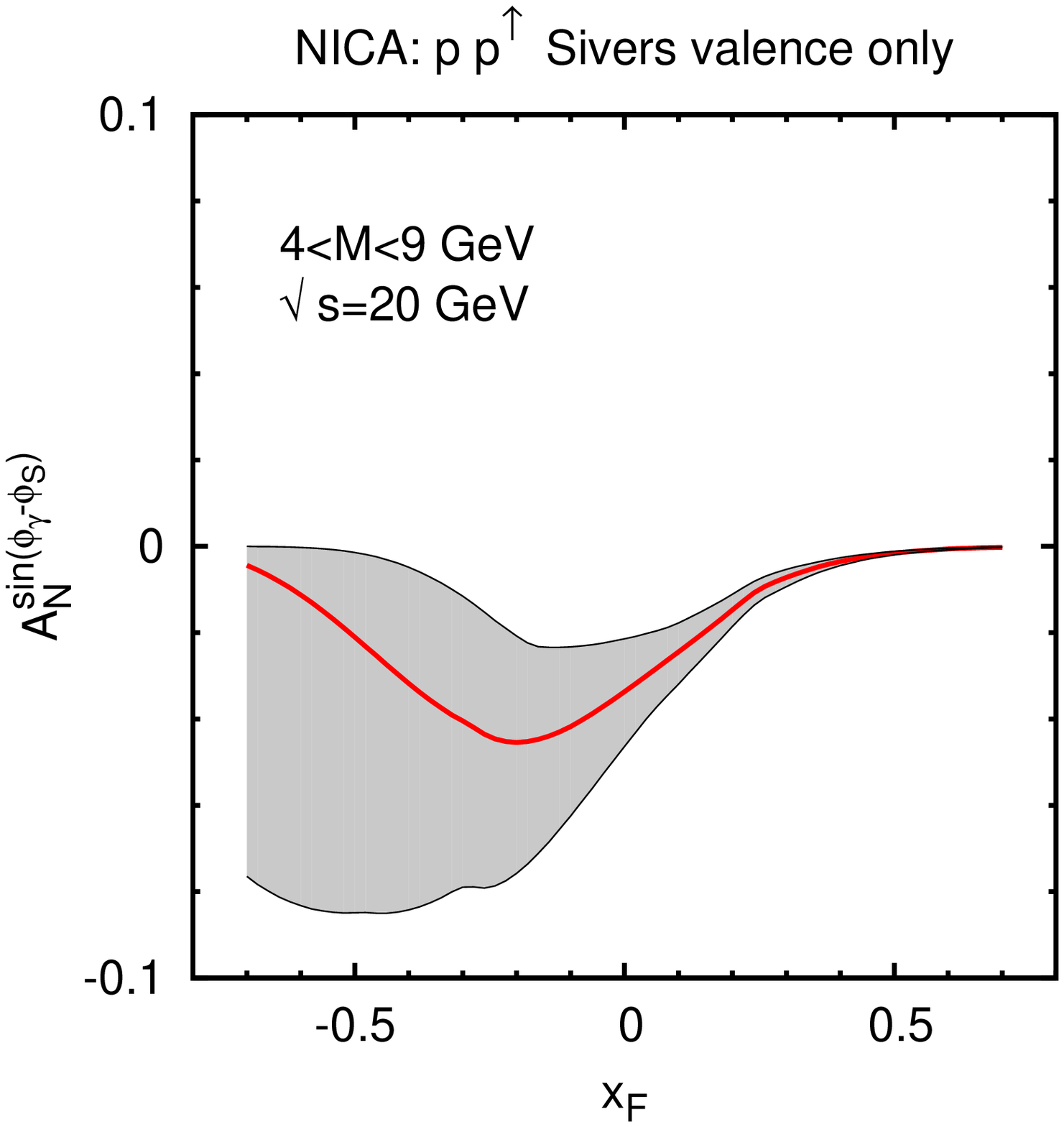}\hspace*{-2cm}
\includegraphics[angle=-90,width=0.45\textwidth]
{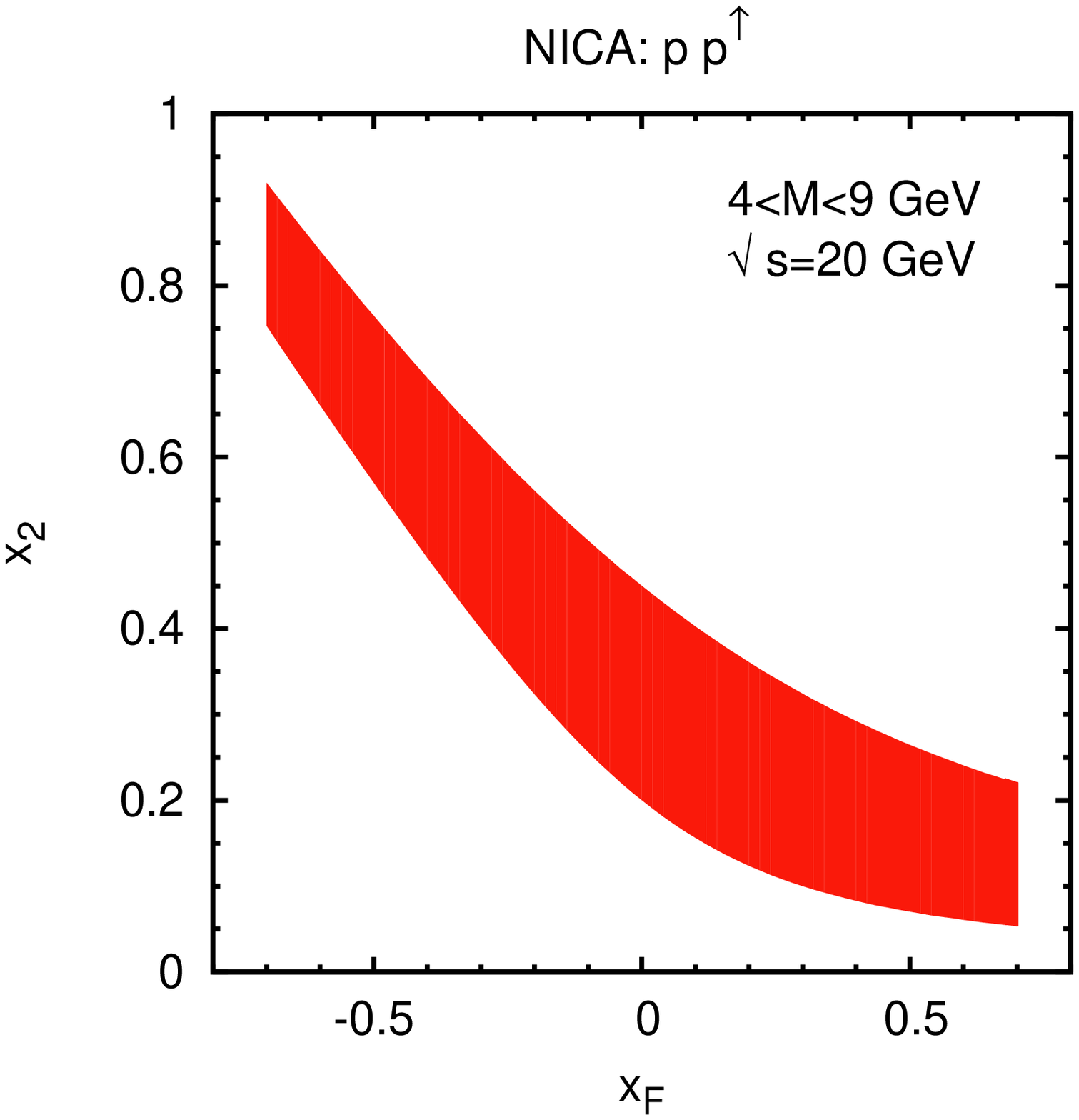}\hspace*{-2cm} \caption{The single spin
asymmetry $A_N^{\sin(\phi_{\gamma}-\phi_S)}$ for the Drell-Yan
process $p\, p^{\uparrow}\to \mu^{+}\mu^{-}\,X$ at JINR (NICA
experiment), as a function of $x_F$.
The integration ranges are $(0 \leq q_T \leq 1)$~GeV and
$(4 \leq M \leq 9)$~GeV. The results are given at $\sqrt s = 20$~GeV;
on the left panel we include the Sivers sea contributions, while on the
central panel we show the results we obtain by including only the
Sivers valence functions. The right panel shows the allowed region
of $x_2$ values as a function of $x_F$.}\label{NICA-AN}
\end{figure}
\end{center}

\begin{center}
\begin{figure}
\includegraphics[angle=-90,width=0.45\textwidth]
{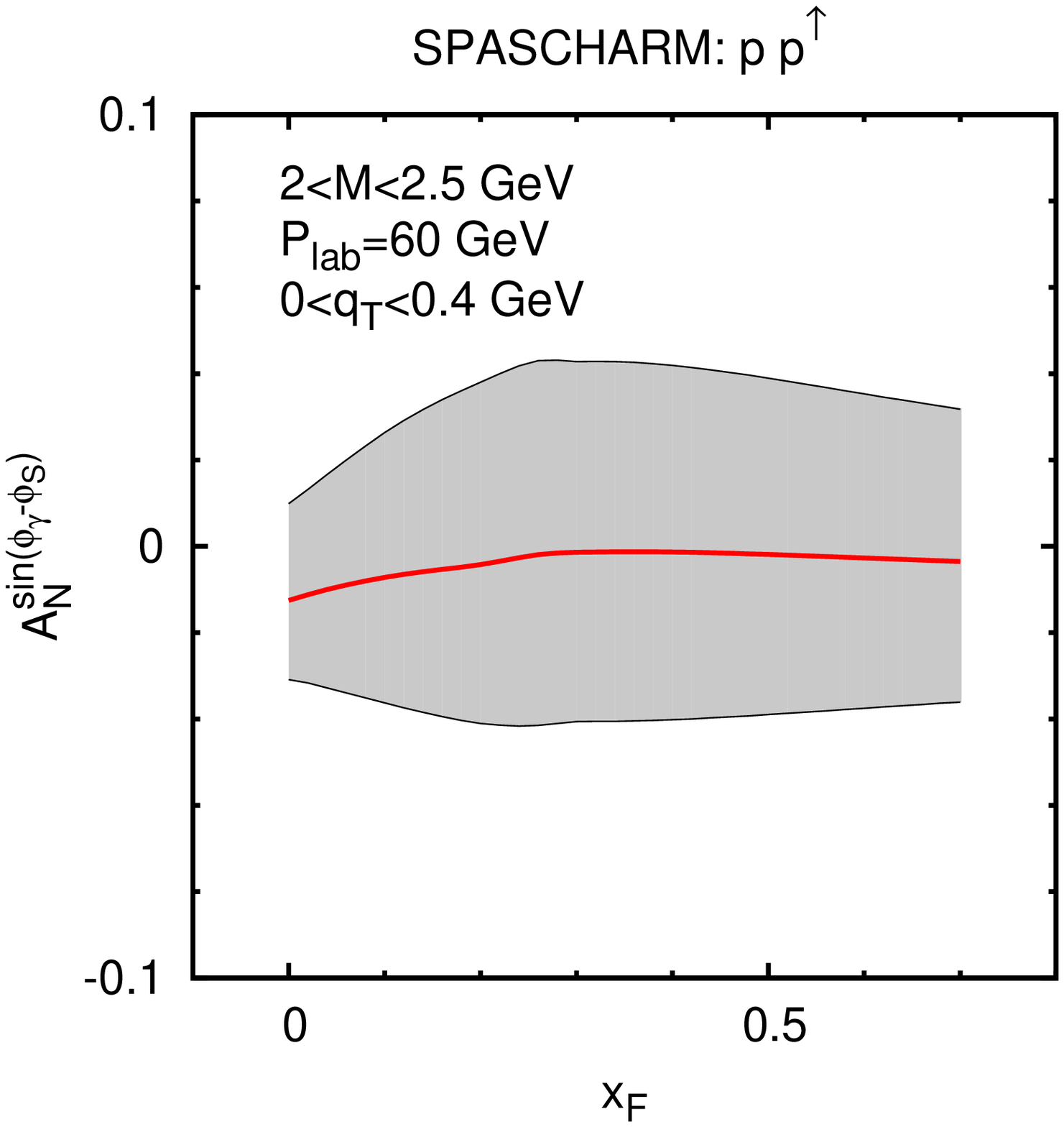}
\includegraphics[angle=-90,width=0.45\textwidth]
{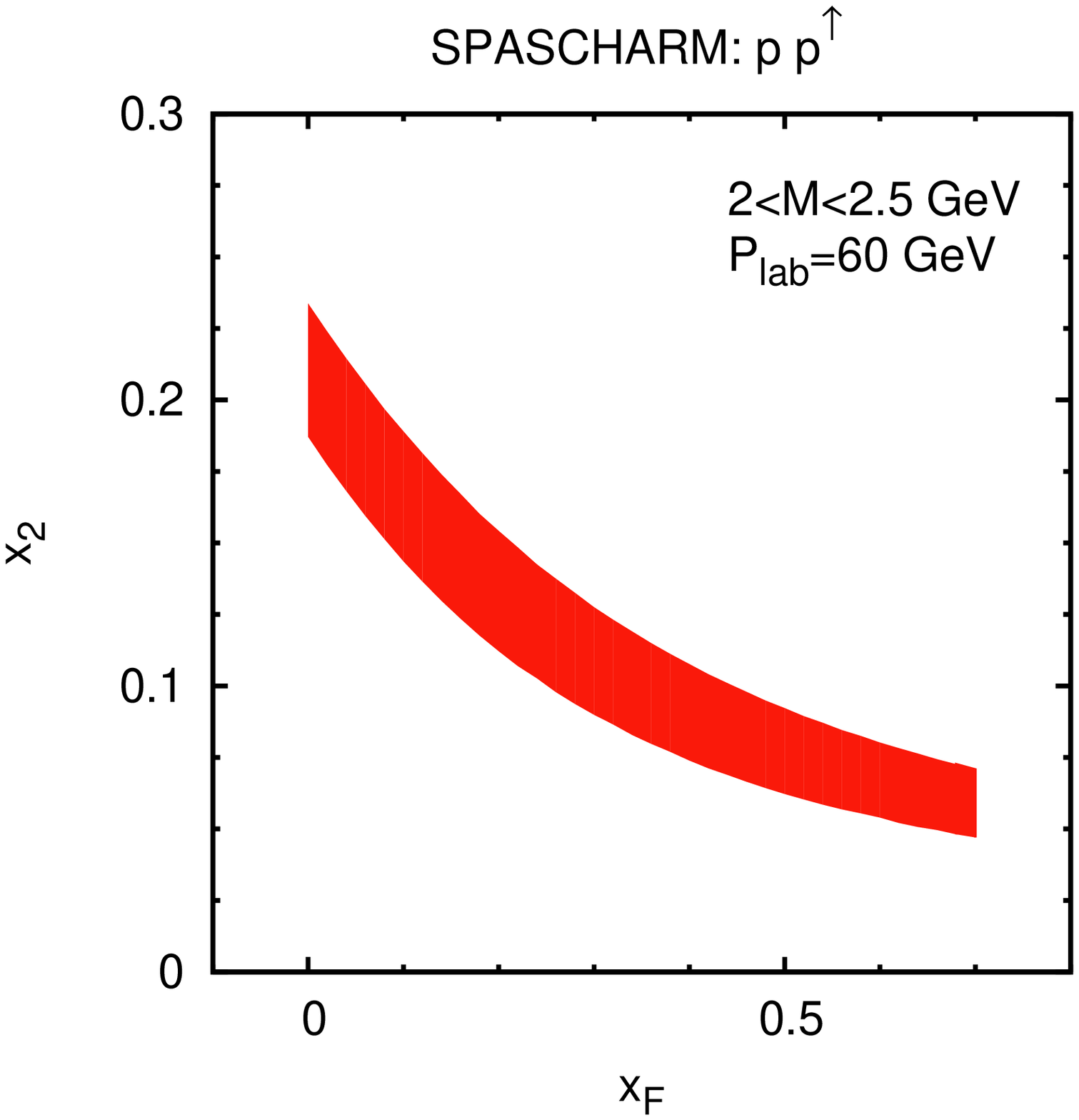}\\
\includegraphics[angle=-90,width=0.45\textwidth]{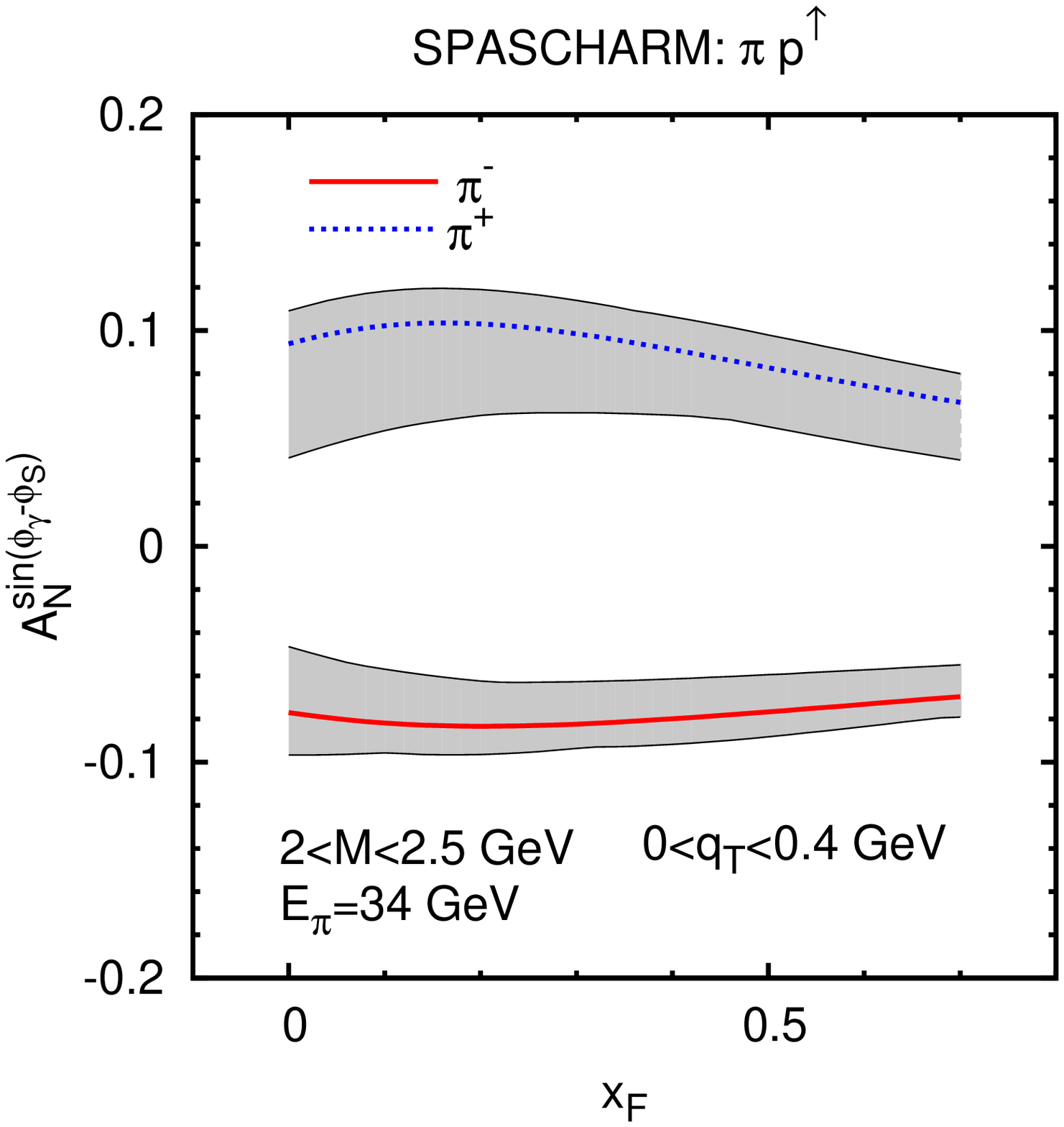}
\includegraphics[angle=-90,width=0.45\textwidth]{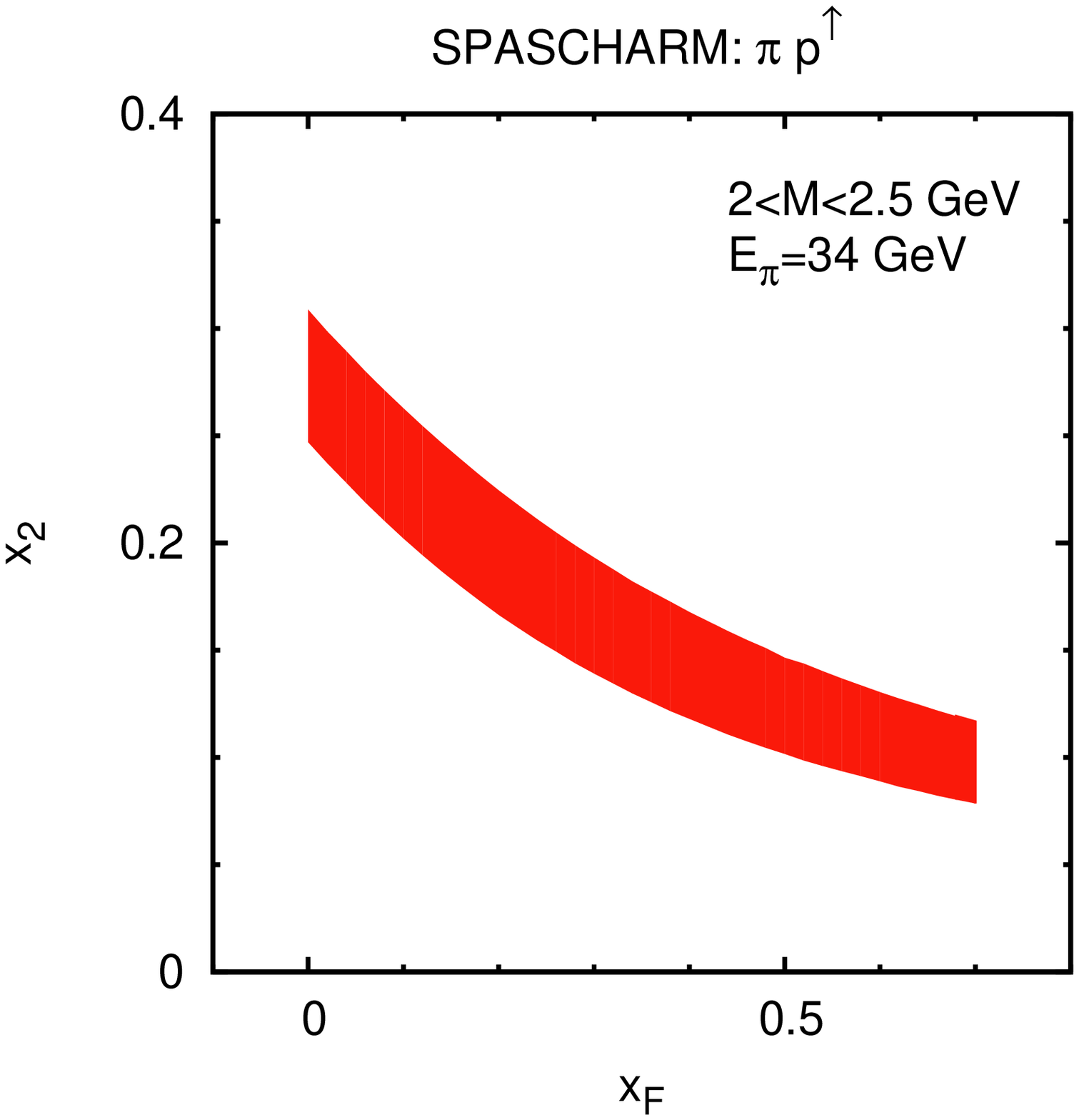}
\caption{The single spin asymmetries
$A_N^{\sin(\phi_{\gamma}-\phi_S)}$ for the Drell-Yan process
$p p^{\uparrow}\to \mu^{+}\mu^{-}\,X$ (upper-left panel) and
$\pi p^{\uparrow}\to \mu^{+}\mu^{-}\,X$ (lower-left panel) at IHEP (SPASCHARM
experiment), as a function of $x_F$. The integration ranges are
$(0 \leq q_T \leq 0.4)$~GeV and $(2 \leq M \leq 2.5)$~GeV. The energies of the
two experiments are, respectively, $P_{\,\rm Lab} = 60$~GeV, corresponding
to $\sqrt{s} = 10.7$~GeV, and $E_\pi = 34$~GeV, corresponding to $\sqrt{s}
= 8$~GeV. The right panels show the allowed region of $x_2$ values as a
function of $x_F$.}\label{PRO-AN}
\end{figure}
\end{center}

Our estimates are obtained using, for the proton, the partonic distribution
functions (PDF) of Ref.~\cite{Gluck:1998xa} (GRV98LO, consistently with the
choice adopted in Ref.~\cite{Anselmino:2008sg}) and, for the pion, those of
Ref.~\cite{Gluck:1991ng}; all the PDFs are evolved (at LO) at the $Q^2 = M^2$ scale of
interest. We have checked that a different choice for the PDFs
of the pion~\cite{Sutton:1991ay} yields very similar results for the $\pi^-$
and results in agreement with those shown here, within the uncertainty bands,
for the $\pi^+$.
Our results are presented in Figs.~\ref{fig:sivers}-\ref{PRO-AN}. The
shaded bands reflect the statistical uncertainty in the extracted values of
the Sivers functions, as explained in Ref.~\cite{Anselmino:2008sg}.
Let us comment on each figure.

\begin{itemize}

\item
{\bf Fig.~\ref{fig:sivers}}, Sivers functions.

For completeness, the Sivers functions used in this analysis are presented.
The sign has been reversed with respect to the Sivers functions obtained
from SIDIS data in Ref.~\cite{Anselmino:2008sg}, according to the findings
of Refs.~\cite{Brodsky:2002rv,Collins:2002kn}:
\be
\Delta^N \! f_{\rm DY} = - \Delta^N \! f_{\rm SIDIS}\; . \label{chsign}
\ee
The first moment of the Sivers function, left panel, is defined as:
\be
\Delta^N \! f_{q/\pup}^{(1)}(x) \equiv \int d^2 \, \bfk_\perp \,\frac{\kt}{4m_p}
\, \Delta^N \! f_{q/\pup}(x, \kt) = - f_{1T}^{\perp (1) q}(x) \>. \label{siv-mom}
\ee
This quantity is, up to a constant resulting from the $d^2 \, \bfk_\perp$
integration, the same as the function $\Delta^N\! f_{q/A^{\uparrow}}(x_1)$
appearing in Eqs.~(\ref{sivfac}),~(\ref{num}).

\item
{\bf Fig.~\ref{Compass-AN}}, COMPASS.

$A_N^{\sin(\phi_{\gamma}-\phi_S)}$ is shown as a function of
$x_F = x_{1} - x_{2}$, left plot, and of $M$, central plot, for the COMPASS
planned experiment, $\pi^\pm \, p^\uparrow \to \mu^+\mu^- X$.
We have integrated both the numerator and denominator of Eq.~(\ref{an-gauss})
over $q_T$ in the range $(0 \leq q_T \leq 1)$ GeV, which is within the region
of validity of our approach. The other integration ranges are
$(4 \leq M \leq 9)$~GeV, at fixed $x_F$, for the left plot and
$0.2 \leq x_F \leq 0.5$, at fixed $M$, for the central plot. The pion beam
energy, in the laboratory frame, is taken to be 160 GeV, corresponding to
$\sqrt s = 17.4$ GeV.

In the right panel we show the kinematical region covered by the COMPASS
experiment, {\it i.e.}~the range of allowed $x_2$ values, which refer to the
polarized proton distributions, as a function of $x_F$. It is interesting to
notice that positive values of $x_F$ correspond to (or at least overlap with)
the $x$ region explored by the SIDIS experiments ($x \leq 0.35$) and the data
used to extract our Sivers functions. Instead, negative values of $x_F$
correspond to larger $x_2$ values, a region yet unexplored by other experiments,
so that our estimates are based on the assumed functional form of the Sivers
function --- not constrained, in that region, by any SIDIS data --- and are bound
to have much larger uncertainties.

Finally, it is important to remark that, as the COMPASS experiment will involve
charged pion beams, $\pi^-(\bar{u}d)$ and $\pi^+(u\bar{d})$, the dominant
elementary process contributing to the asymmetry will be
$\bar u _{\pi^-} u_p \to \mu^+\mu^-$ for the $\pi^-$ beam and
$\bar d _{\pi^+} d_p\to \mu^+\mu^-$ for the $\pi^+$ beam. Consequently, the
prediction of the sign change of the Sivers function in SIDIS and Drell-Yan
processes~\cite{Brodsky:2002rv,Collins:2002kn} can be clearly tested, as the
sign of the asymmetry for $\pi^-$ ($\pi^+$) beam is given by the sign of
the $u$ ($d$) Sivers function, which is well established.

\item
{\bf Fig.~\ref{Compass-AN-low}}, COMPASS (low $M$).

We show our estimates for the single spin asymmetry and the corresponding
allowed kinematical region in the $x_2-x_F$ plane, for COMPASS experiments
performed at lower $M$ values, $(2.0\leq M \leq 2.5)$~GeV and in a more
limited $q_T$ region, $(0\leq q_T \leq 0.4)$~GeV (so that the constraints
of Eq.~(\ref{kinr}) remain true). This exploits the possibility of DY
measurements in the low mass region, below the $J/\psi$ and above the $\rho$
resonance peaks, where data should be much more abundant.

\item
{\bf Fig.~\ref{RHIC-AN}}, RHIC.

$A_N^{\sin(\phi_{\gamma}-\phi_S)}$ is shown as a function of  $x_F$ and $M$
for RHIC experiments, $p^\uparrow p \to \ell^+\ell^- X$, at
$\sqrt s = 200$~GeV. The integration ranges for $q_T$ and $M$ are the same
as in Fig.~\ref{Compass-AN}, with the further constraint $0 \leq y \leq3$,
according to the experimental kinematical cuts. The right panel shows
the kinematical region of RHIC for $x_1$ as a function of $x_F$.
This region covers the range already explored by SIDIS measurements
($x \leq 0.35$) where the Sivers functions are reliably constrained, and
expands into much higher values of $x$. The maximum value of
the asymmetry ($\sim$ 10\%) shown in the left panel is expected at
$x_F\simeq 0.2$ which corresponds (see the right panel) to $x_1\simeq 0.2$,
where the valence Sivers functions reach their maximum, see
Fig.~\ref{fig:sivers}.

The asymmetry for larger values of $x_F$ ($\,> 0.4$) is obtained by using our
extracted Sivers functions at large values of $x$ where SIDIS data offer no
constraint. This reflects into the huge uncertainty band. Let us remind that
in Ref.~\cite{Anselmino:2008sg} we have assumed the $\beta$ parameters, which
determine the large $x$ behaviour of the Sivers functions, to be equal for all
flavours. With such an assumption the asymmetry decreases fast for high values
of $x_F$, although with a large uncertainty. The parameterization of our
previous analysis \cite{Anselmino:2005ea}, where $\beta_u$ and $\beta_d$ were
different for $u$ and $d$ flavours, would lead to a larger asymmetry at high
$x_F$ (see Fig.~6 of Ref.~\cite{Anselmino:2005ea}). Measurements in the region
of high $x_F$ could then offer an opportunity to test the flavour-blind
$\beta$ assumption. Moreover, data in the negative $x_F$ region would test the contribution of the sea Sivers functions, as first pointed out in
Ref.~\cite{Collins:2005rq}.

\item
{\bf Fig.~\ref{PAX-AN}}, PAX.

Estimates are given for the planned GSI-PAX experiment~\cite{Barone:2005pu},
$p^\uparrow \bar p \to \ell^+\ell^- X$, in the asymmetric collider mode at
$\sqrt s = 14.14$~GeV, with a polarized proton beam. This experiment would
have the big advantage of having proton and antiproton beams, thus increasing
the number of Drell-Yan events. $A_N^{\sin(\phi_{\gamma}-\phi_S)}$ is shown
as a function of $x_F$ (integrating over $(4 \leq M \leq 6)$~GeV, with
the constraint $|y|<1$), and as a function of $M$ (integrating over $x_F$
with $|y|<1)$; in both cases $q_T$ is integrated in the range
$(0 \leq q_T \leq 1)$~GeV.

The negative $x_F$ region gets most contribution from the valence $x_1$
region, which is well explored by SIDIS experiments and where the Sivers
distributions are reliably known. In this region the asymmetry is large
(10-15\% in size) and sensitive only to the $u$ and $d$ Sivers functions.
On the contrary, at large and positive $x_F$ values the situation becomes
similar to that described for RHIC, as we need the Sivers functions
in the large $x$-range, which are more poorly known. Once again, this results
in large uncertainties of our estimates.

The PAX collaboration aims at having a beam of polarized antiprotons, which,
in combination with a beam of polarized protons, would allow a unique direct
way of measuring the transversity distributions~\cite{Barone:2005pu}. In such
a case, one could also measure the single spin Sivers asymmetry in the process
$p \, \bar p^\uparrow \to \ell^+\ell^- X$; this is related to the one considered
here by charge conjugation invariance and the relation of Eq.~(\ref{ab-ba}),
\be
A_N^{\sin(\phi_{\gamma} - \phi_S)}(p \, \bar p^\uparrow \to \gamma^* X;
\> x_F, M, q_T) =
- A_N^{\sin(\phi_{\gamma} - \phi_S)}(p^\uparrow \bar p \to \gamma^* X;
\> -x_F, M, q_T) \>.
\label{ab-ba2}
\ee

\item
{\bf Fig.~\ref{PANDA-AN}}, PANDA.

The GSI-PANDA experiment~\cite{Brinkmann:2007zz} will run with an antiproton beam scattering off a
proton target, at the energy $E_{\bar p} =15$~GeV, corresponding
to $\sqrt{s}=5.47$~GeV. It might be worth trying to have a polarized target,
$\bar p \, p^\uparrow \to \ell^+\ell^- X$. We then show predictions for
$A_N^{\sin(\phi_{\gamma}-\phi_S)}$ as a function of $x_F$; the integration
ranges, due to the moderate energy, are $(2.0\leq M\leq 2.5)$~GeV and
$(0\leq q_T \leq 0.4)$~GeV. The kinematical range which could be explored is
entirely overlapping with that covered by the SIDIS data, so that this
experiment would supply a perfect consistency check of the HERMES and COMPASS
results, and of the crucial Eq.~(\ref{chsign}). The asymmetry turns out
to be sizable and well definite in sign over the full kinematical range.

\item
{\bf Fig.~\ref{JPARC-AN}}, J-PARC.

J-PARC might measure Drell-Yan single spin asymmetries $A_N^{\sin(\phi_{\gamma}-\phi_S)}$
generated in polarized proton-proton collisions at $E_p=50$~GeV~\cite{jparc-loi15}, corresponding
to $\sqrt{s}=9.78$~GeV (see the figure caption for further information).
As shown on the right panel, the kinematical range covered by J-PARC would be
almost entirely complementary to that explored in the SIDIS experiments. These measurements could then be of great importance to obtain information on the
large-$x$ behaviour of the Sivers distribution functions.

\item
{\bf Fig.~\ref{NICA-AN}}, NICA.

NICA is a planned Drell-Yan experiment to be performed at JINR in Dubna~\cite{Trubnikov:2008zz}.
We present our estimates for an unpolarized proton beam scattering off
polarized protons at $\sqrt{s}=20$~GeV. As shown in the right panel, the
region of large and negative $x_F$ ($\,< -0.5$) corresponds to the range of
$x_2$ values for which we have no information from SIDIS experiments. Therefore
the measurements of $A_N^{\sin(\phi_{\gamma}-\phi_S)}$ in this region would
provide information on the large $x$ behaviour of the Sivers functions.
Our results have smaller uncertainties in the region $-0.4 \lesssim x_F
\lesssim 0.1$ which corresponds to the $x_2$ valence region
and where the asymmetry is related to the convolution
$4\, f_{\bar{u}} \otimes \Delta^N\!f_u + f_{\bar{d}} \otimes \Delta^N\!f_d$.
The comparison between the left and the central plots shows that the positive
$x_F$ region is dominated by the Sivers sea contribution. More information can
be found in the figure caption.

\item
{\bf Fig.~\ref{PRO-AN}}, SPASCHARM.

The SPASCHARM experiment~\cite{Vasiliev:2007nz} at IHEP in Protvino (Russia) plans to measure Drell-Yan
processes both in polarized proton-proton scattering at $P_{\,\rm Lab}=60$~GeV,
corresponding to $\sqrt{s} = 10.7$~GeV (upper panels) and in pion-proton
scattering at $E_\pi = 34$~GeV, corresponding to $\sqrt{s}
= 8$~GeV (lower panels). SPASCHARM would mainly collect data in the
range of low values of $M$ [($2.0 \leq M \leq 2.5)$~GeV] and in the valence
region of the Sivers functions (see the figure caption for more details).
Therefore, SPASCHARM measurements of pion-proton asymmetries, like the analogous
COMPASS case, could provide a stringent test of the sign change of the Sivers
function in Drell-Yan and SIDIS processes.
\end{itemize}

\section{Comments and conclusions}

We have given clear and simple estimates of Sivers SSAs in Drell-Yan processes,
which could be measured in experiments which are either in preparation or being
planned. Our results show large values of these asymmetries, which, despite
the uncertainty in magnitude related to the extraction of the Sivers functions
from SIDIS data, have an unambiguous sign. This sign originates from the current
understanding of the origin of SSAs at the partonic level \cite{Collins:2002kn}
which predicts that the Sivers distributions must enter with opposite signs in
SIDIS and Drell-Yan processes. Then, the mere measurement of this sign would,
alone, provide a stringent and important test of the present interpretation of
the observed SSAs in terms of TMD distribution functions, coupled to
initial or final state QCD interactions.

This test, if successful, would be a significant and decisive step
towards a basic and consistent description of subtle phenomena like
the challenging SSAs in terms of elementary pQCD dynamics and
fundamental properties of the nucleon structure involving quark
intrinsic motion. Drell-Yan processes at small $q_T$ values,
exploring the quark content of the nucleon without the complications
of fragmentation processes, might play a crucial role in the future
exploration of the proton structure.

\section*{Acknowledgements}

We acknowledge support of the European Community - Research Infrastructure
Activity under the FP6 ``Structuring the European Research Area''
program (HadronPhysics, contract number RII3-CT-2004-506078).
M.A.,~M.B., and A.P.~acknowledge partial support by MIUR under Cofinanziamento
PRIN 2006. This work is partially supported by the Helmholtz Association through
funds provided to the virtual institute ``Spin and strong QCD''(VH-VI-231).


\begin{thebibliography}{45}
\expandafter\ifx\csname natexlab\endcsname\relax\def\natexlab#1{#1}\fi
\providecommand{\enquote}[1]{``#1''}
\expandafter\ifx\csname url\endcsname\relax
  \def\url#1{\texttt{#1}}\fi
\expandafter\ifx\csname urlprefix\endcsname\relax\def\urlprefix{URL }\fi
\providecommand{\eprint}[2][]{\url{#2}}

\bibitem[Anselmino et~al.(2008)]{Anselmino:2008sg}
M.~Anselmino, et~al., \emph{Eur. Phys. J.} \textbf{A39}, 89 (2009),
  \eprint{arXiv:0805.2677 [hep-ph]}.

\bibitem[Kane et~al.(1978)]{Kane:1978nd}
G.~L. Kane, J.~Pumplin, and W.~Repko, \emph{Phys. Rev. Lett.} \textbf{41}, 1689
  (1978).

\bibitem[Sivers(1990)]{Sivers:1989cc}
D.~W. Sivers, \emph{Phys. Rev.} \textbf{D41}, 83 (1990).

\bibitem[Sivers(1991)]{Sivers:1990fh}
D.~W. Sivers, \emph{Phys. Rev.} \textbf{D43}, 261 (1991).

\bibitem[Bacchetta et~al.(2004)]{Bacchetta:2004jz}
A.~Bacchetta, U.~D'Alesio, M.~Diehl, and C.~A. Miller, \emph{Phys. Rev.}
  \textbf{D70}, 117504 (2004), \eprint{arXiv:hep-ph/0410050}.

\bibitem[Sivers(2006)]{Sivers:2006rg}
D.~Sivers, \emph{Phys. Rev.} \textbf{D74}, 094008 (2006),
  \eprint{arXiv:hep-ph/0609080}.

\bibitem[Burkardt and Schnell(2006)]{Burkardt:2005km}
M.~Burkardt, and G.~Schnell, \emph{Phys. Rev.} \textbf{D74}, 013002 (2006),
  \eprint{arXiv:hep-ph/0510249}.

\bibitem[Airapetian et~al.(2005)]{Airapetian:2004tw}
A.~Airapetian, et~al.~(HERMES), \emph{Phys. Rev. Lett.} \textbf{94}, 012002 (2005),
  \eprint{arXiv:hep-ex/0408013}.

\bibitem[Diefenthaler(2007)]{Diefenthaler:2007rj}
M.~Diefenthaler (HERMES), \eprint{arXiv:0706.2242 [hep-ex]}.

\bibitem[Alexakhin et~al.(2005)]{Alexakhin:2005iw}
V.~Y. Alexakhin, et~al.~(COMPASS), \emph{Phys. Rev. Lett.} \textbf{94}, 202002 (2005),
  \eprint{arXiv:hep-ex/0503002}.

\bibitem[Martin(2006)]{Martin:2007au}
A.~Martin (COMPASS), \emph{Czech. J. Phys.} \textbf{56}, F33 (2006),
  \eprint{arXiv:hep-ex/0702002}.

\bibitem[Alekseev et~al.(2008)]{Alekseev:2008dn}
M.~Alekseev, et~al.~(COMPASS), \eprint{arXiv:0802.2160 [hep-ex]}.

\bibitem[Anselmino et~al.(2005{\natexlab{a}})]{Anselmino:2005nn}
M.~Anselmino, et~al., \emph{Phys. Rev.} \textbf{D71}, 074006
  (2005{\natexlab{a}}), \eprint{arXiv:hep-ph/0501196}.

\bibitem[Anselmino et~al.(2005{\natexlab{b}})]{Anselmino:2005ea}
M.~Anselmino, et~al., \emph{Phys. Rev.} \textbf{D72}, 094007
  (2005{\natexlab{b}}), \eprint{arXiv:hep-ph/0507181}.

\bibitem[Vogelsang and Yuan(2005)]{Vogelsang:2005cs}
W.~Vogelsang, and F.~Yuan, \emph{Phys. Rev.} \textbf{D72}, 054028 (2005),
  \eprint{arXiv:hep-ph/0507266}.

\bibitem[Collins et~al.(2006{\natexlab{a}})]{Collins:2005ie}
J.~C. Collins, et~al., \emph{Phys. Rev.} \textbf{D73}, 014021
  (2006{\natexlab{a}}), \eprint{arXiv:hep-ph/0509076}.

\bibitem[Adams et~al.(1991{\natexlab{a}})]{Adams:1991rx}
D.~L. Adams, et~al.~(E581) \emph{Phys. Lett.} \textbf{B261}, 197
  (1991{\natexlab{a}}).

\bibitem[Adams et~al.(1991{\natexlab{b}})]{Adams:1991cs}
D.~L. Adams, et~al.~(E704) \emph{Phys. Lett.} \textbf{B264}, 462
  (1991{\natexlab{b}}).

\bibitem[Adams et~al.(1995)]{Adams:1995gg}
D.~L. Adams, et~al.~(E704) \emph{Phys. Lett.} \textbf{B345}, 569 (1995).

\bibitem[Adams et~al.(2004)]{Adams:2003fx}
J.~Adams, et~al.~(STAR) \emph{Phys. Rev. Lett.} \textbf{92}, 171801 (2004),
  \eprint{arXiv:hep-ex/0310058}.

\bibitem[Arsene et~al.(2008)]{:2008mi}
I.~Arsene, et~al.~(BRAHMS) \emph{Phys. Rev. Lett.} \textbf{101}, 042001 (2008),
  \eprint{arXiv:0801.1078 [nucl-ex]}.

\bibitem[Anselmino et~al.(1995)]{Anselmino:1994tv}
M.~Anselmino, M.~Boglione, and F.~Murgia, \emph{Phys. Lett.} \textbf{B362},
  164 (1995), \eprint{arXiv:hep-ph/9503290}.

\bibitem[Anselmino and Murgia(1998)]{Anselmino:1998yz}
M.~Anselmino, and F.~Murgia, \emph{Phys. Lett.} \textbf{B442}, 470 (1998),
  \eprint{arXiv:hep-ph/9808426}.

\bibitem[D'Alesio and Murgia(2004)]{D'Alesio:2004up}
U.~D'Alesio, and F.~Murgia, \emph{Phys. Rev.} \textbf{D70}, 074009 (2004),
  \eprint{arXiv:hep-ph/0408092}.

\bibitem[Levorato(2008)]{Levorato:2008tv}
S.~Levorato (COMPASS), \eprint{arXiv:0808.0086 [hep-ex]}.

\bibitem[Collins(1993)]{Collins:1992kk}
J.~C. Collins, \emph{Nucl. Phys.} \textbf{B396}, 161 (1993).

\bibitem[Anselmino et~al.(2002)]{Anselmino:2001vn}
M.~Anselmino, V.~Barone, A.~Drago, and F.~Murgia, \emph{Nucl. Phys. Proc.
  Suppl.} \textbf{105}, 132 (2002), \eprint{arXiv:hep-ph/0111044}.

\bibitem[Brodsky et~al.(2002{\natexlab{a}})]{Brodsky:2002cx}
S.~J. Brodsky, D.~S. Hwang, and I.~Schmidt, \emph{Phys. Lett.} \textbf{B530},
  99 (2002{\natexlab{a}}), \eprint{arXiv:hep-ph/0201296}.

\bibitem[Collins(2002)]{Collins:2002kn}
J.~C. Collins, \emph{Phys. Lett.} \textbf{B536}, 43 (2002),
  \eprint{arXiv:hep-ph/0204004}.

\bibitem[Brodsky et~al.(2002{\natexlab{b}})]{Brodsky:2002rv}
S.~J. Brodsky, D.~S. Hwang, and I.~Schmidt, \emph{Nucl. Phys.} \textbf{B642},
  344 (2002{\natexlab{b}}), \eprint{arXiv:hep-ph/0206259}.

\bibitem[Anselmino et~al.(2003)]{Anselmino:2002pd}
M.~Anselmino, U.~D'Alesio, and F.~Murgia, \emph{Phys. Rev.} \textbf{D67},
  074010 (2003), \eprint{arXiv:hep-ph/0210371}.

\bibitem[Collins et~al.(2006{\natexlab{b}})]{Collins:2005rq}
J.~C. Collins, et~al., \emph{Phys. Rev.} \textbf{D73}, 094023
  (2006{\natexlab{b}}), \eprint{arXiv:hep-ph/0511272}.

\bibitem[D'Alesio and Murgia(2008)]{D'Alesio:2007jt}
U.~D'Alesio, and F.~Murgia, \emph{Prog. Part. Nucl. Phys. 61} \textbf{2008},
  394 (2008), \eprint{arXiv:0712.4328 [hep-ph]}.

\bibitem[Arnold et~al.(2008)]{Arnold:2008kf}
S.~Arnold, A.~Metz, and M.~Schlegel  (2008), \eprint{arXiv:0809.2262 [hep-ph]}.

\bibitem[Collins et~al.(1985)]{Collins:1984kg}
J.~C. Collins, D.~E. Soper, and G.~Sterman, \emph{Nucl. Phys.} \textbf{B250},
  199 (1985).

\bibitem[Ji et~al.(2005)]{Ji:2004wu}
X.-d. Ji, J.-p. Ma, and F.~Yuan, \emph{Phys. Rev.} \textbf{D71}, 034005 (2005),
  \eprint{arXiv:hep-ph/0404183}.

\bibitem[Ji et~al.(2004)]{Ji:2004xq}
X.-d. Ji, J.-P. Ma, and F.~Yuan, \emph{Phys. Lett.} \textbf{B597}, 299
  (2004), \eprint{arXiv:hep-ph/0405085}.

\bibitem[Ji et~al.(2006)]{Ji:2006vf}
X.~Ji, J.-w. Qiu, W.~Vogelsang, and F.~Yuan, \emph{Phys. Rev.} \textbf{D73},
  094017 (2006), \eprint{arXiv:hep-ph/0604023}.

\bibitem[Collins and Soper(1977)]{Collins:1977iv}
J.~C. Collins, and D.~E. Soper, \emph{Phys. Rev.} \textbf{D16}, 2219 (1977).

\bibitem[Efremov et~al.(2005)]{Efremov:2004tp}
A.~V. Efremov, K.~Goeke, S.~Menzel, A.~Metz, and P.~Schweitzer, \emph{Phys.
  Lett.} \textbf{B612}, 233 (2005), \eprint{arXiv:hep-ph/0412353}.

\bibitem[Gluck et~al.(1998)]{Gluck:1998xa}
M.~Gluck, E.~Reya, and A.~Vogt, \emph{Eur. Phys. J.} \textbf{C5}, 461
  (1998), \eprint{arXiv:hep-ph/9806404}.

\bibitem[Gluck et~al.(1992)]{Gluck:1991ng}
M.~Gluck, E.~Reya, and A.~Vogt, \emph{Z. Phys.} \textbf{C53}, 127 (1992).

\bibitem[Sutton et~al.(1992)]{Sutton:1991ay}
P.~J. Sutton, A.~D. Martin, R.~G. Roberts, and W.~J. Stirling, \emph{Phys.
  Rev.} \textbf{D45}, 2349 (1992).

\bibitem[Barone et~al.(2005)]{Barone:2005pu}
V.~Barone, et~al.~(PAX), \eprint{arXiv:hep-ex/0505054}.

\bibitem{Brinkmann:2007zz}
K.~T.~Brinkmann~(PANDA), \emph{Nucl. Phys.} \textbf{A790}, 75 (2007).

\bibitem{jparc-loi15}
D.~Dutta, et~al., J-PARC Letter of Intent L15, URL:~http://psux1.kek.jp/jhf-np/LOIlist/LOIlist.html

\bibitem{Trubnikov:2008zz}
G.~V.~Trubnikov, et~al.~(NICA), \emph{11th European Particle Accelerator Conference (EPAC 08)},
Genoa, Italy, 23-27 June 2008, URL:~http://accelconf.web.cern.ch/AccelConf/e08/papers/wepp029.pdf

\bibitem[Vasiliev et~al.(Dubna, Russia, September 3-7, 2007)]{Vasiliev:2007nz}
A.~N. Vasiliev, et~al., \emph{Proceedings of the XII Workshop on High Energy
  Spin Physics, DSPIN-07}, p.~368 (Dubna, Russia, September 3-7, 2007),
  \eprint{arXiv:hep-ex/0712.2691}.

\end{thebibliography}

%
%
%
%

%

%
\end{document}